\documentclass[manuscript,screen]{acmart}

\acmJournal{TOSEM}
\usepackage{amsmath}
\usepackage{booktabs}
\usepackage{tabularx}
\usepackage{array}
\usepackage{multirow}
\usepackage{threeparttable}
\usepackage{graphicx}
\usepackage{listings}
\usepackage{pifont}
\usepackage{enumitem}
\usepackage{algorithm}
\usepackage{algpseudocode}
\usepackage{appendix}

\newcommand{\code}[1]{\texttt{#1}}
\newcommand{\etal}{et~al.}
\newcolumntype{Y}{>{\raggedright\arraybackslash}X}

\title{Which Code Changes Need Re-embedding? CIPBench and the Limits of Multi-Signal Index Maintenance}

\author{Nannan Zhao}
\affiliation{%
  \institution{Northwestern Polytechnical University}
  \city{Xi'an}
  \country{China}
}
\email{nannanzhao@nwpu.edu.cn}

\author{Jiameng Zhang}
\affiliation{%
  \institution{Northwestern Polytechnical University}
  \city{Xi'an}
  \country{China}
}
\email{zhangjiameng@mail.nwpu.edu.cn}

\author{Taoyu Zhong}
\affiliation{%
  \institution{Northwestern Polytechnical University}
  \city{Xi'an}
  \country{China}
}
\email{zhongtaoyu@mail.nwpu.edu.cn}

\author{Hao Wang}
\affiliation{%
  \institution{Northwestern Polytechnical University}
  \city{Xi'an}
  \country{China}
}
\email{kingshulk989@mail.nwpu.edu.cn}

\author{Bo Wang}
\affiliation{%
  \institution{Northwestern Polytechnical University}
  \city{Xi'an}
  \country{China}
}
\email{wb2022302586@mail.nwpu.edu.cn}

\begin{document}

\begin{abstract}
Repository-level retrieval-augmented generation systems must keep vector indexes synchronized with continuously evolving code, yet there is no principled basis for deciding which changed code units actually need new embeddings. We ask whether the retrieval impact of a code change can be anticipated before that embedding is computed.

Across 10 open-source repositories in Python, Go, and JavaScript, we characterize how a change propagates from syntactic substantiality through embedding displacement to retrieval impact. The links are informative but not deterministic: syntactic class separates displacement sharply, yet impact depends on a unit's retrieval exposure as much as on how far its representation moves, and recurring counterexamples bound what any single signal can decide. We formalize the decision as \textbf{CIPBench}, a benchmark that scores maintenance policies under a strict pre-embedding information boundary.

Prediction works, and one signal exhausts it. A ranker using only a unit's change history substantially reduces retrieval harm relative to random and recency-based scheduling. Combining it with three further signals — individually discriminative and largely unassociated with one another — leaves the decision statistically unchanged across every budget we evaluate, for reasons we trace to the structure of the score rather than to its tuning. What does pay off is not ranking but filtering: units whose text is unchanged from the indexed state can be skipped with no risk and, in the most reversion-heavy project, at nearly half the embedding cost.
Signal-level informativeness is therefore not evidence of decision-level value, and separating the two requires evaluating the decision a signal produces rather than its association with the outcome.
\end{abstract}


% % ACM CCS Concepts
% \ccsdesc[500]{Software and its engineering~Software evolution}
% \ccsdesc[300]{Information systems~Search engine indexing}
% \ccsdesc[300]{Information systems~Retrieval models and ranking}



\maketitle

\section{Introduction}
\label{sec:intro}

Modern software repositories evolve continuously, while repository-level
retrieval-augmented generation (RAG) systems rely on vector indexes built from
particular repository snapshots~\cite{zhang2023repocoder,wang2025coderagbench}. As code
changes, these indexes become stale: a modified function may no longer be represented by
the vector stored in the index, causing the retriever to return outdated context to the
downstream large language model. Keeping the index synchronized with the repository is
therefore a recurring maintenance task in repository-level code intelligence systems.

The straightforward solution is to re-embed every changed code unit, but at repository
scale this wastes computation. Full re-embedding scales with corpus size, while
per-change incremental re-embedding still recomputes a representation for every observed
update---including changes with little or no effect on retrieval. The shape of the
problem is familiar from predictive test selection, where a change arrives, a budget
constrains the downstream work it triggers, and a signal derived from the change must
decide what to spend that budget
on~\cite{machalica2019predictive,rothermel1997safe}. Here the downstream work is
embedding, and the decision is:
\emph{when a code unit changes, does its representation actually need to be recomputed?}

Answering this is less straightforward than it appears. A natural intuition is that the
syntactic nature of a change should serve as a signal: a formatting or documentation edit
may be less consequential than a modification to a function body or signature. Prior work
already cautions against taking that correspondence for granted---Fluri and Gall showed
that textual edits need not correspond to substantive program
changes~\cite{fluri2006change}, and Kawrykow and Robillard documented how much of a
version history consists of non-essential
modifications~\cite{kawrykow2011nonessential}---but the gap that matters here opens one
layer further downstream. The maintenance objective is not syntactic substantiality but
whether a change alters the unit's representation enough to affect retrieval, and these
need not coincide. A syntactically substantial edit may leave the embedding almost
unchanged; an embedding may move substantially without altering any relevant top-$K$
result; and a small movement may change retrieval when the unit lies near a ranking
boundary. \textit{The question is therefore not whether code has changed, but how a change
propagates from syntax, through embedding space, to retrieval outcomes}.

Existing research supplies pieces of this problem without characterizing the propagation
end to end. Fine-grained code-change analysis---ChangeDistiller~\cite{fluri2007changedistilling},
GumTree~\cite{falleri2014gumtree}, non-essential-change
detection~\cite{kawrykow2011nonessential}---describes what changed and relates it to
defects, tests, refactorings, or affected program entities, but never to embedding
displacement or retrieval impact. Approaching from the other side, code-search benchmarks
such as CodeSearchNet~\cite{husain2019codesearchnet} and CoSQA~\cite{huang2021cosqa} and
RAG evaluators such as RAGAS~\cite{es2024ragas} and ARES~\cite{saadfalcon2024ares} assume
a fixed indexed corpus, while dynamic vector-index systems such as
SPFresh~\cite{xu2023spfresh}, Ada-IVF~\cite{mohoney2024adaivf} and
Quake~\cite{mohoney2025quake} begin \emph{after} the new vector exists, optimizing how it
is incorporated rather than whether computing it was necessary. Drift-aware retrieval
such as DeDrift~\cite{baranchuk2023dedrift} likewise responds to drift once it has
occurred. This leaves an upstream decision unevaluated: which changed units should be
re-embedded at all?

We study that decision as an empirical and evaluative problem, characterizing a
three-layer funnel from \emph{syntactic substantiality} through \emph{embedding
displacement} to \emph{retrieval impact} across 10 open-source repositories in Python, Go
and JavaScript---2,366 function-level units and 8,936 real updates---with key
measurements replicated over three embedding models. The layers turn out to be related
but not interchangeable. Change type separates displacement sharply, yet displacement
alone does not determine retrieval: the effect is jointly moderated by a unit's position
in the retrieval space. Recurring counterexamples---documentation edits that move an
embedding without moving retrieval, comment edits that remain retrieval-relevant despite
leaving executable syntax untouched, reversions that move nothing at all---mark the
boundaries of any single-signal rule. Change history supplies a further inexpensive
prior, following the persistence that G\^irba \etal\ observed for change-prone
entities~\cite{girba2004yesterday}.

These findings expose a second problem: no standard way exists to evaluate the
re-embedding decision itself. A code-search benchmark asks whether a system retrieves the
right code from a snapshot. An index-maintenance benchmark must ask whether a system can
decide, \emph{before observing the updated embedding or its retrieval effect}, which
changes deserve recomputation. We therefore introduce \textbf{CIPBench}(open-source code available at \url{https://github.com/nnzhaocs/CIPBench}), which turns
evolving-index maintenance into a reproducible decision problem under a strict
pre-embedding information boundary, with two tracks: filtering, which exposes the
safety--savings trade-off when changed units are skipped, and budget-constrained
scheduling, which evaluates which units to prioritize when only a fraction can be
re-embedded.

Finally, we ask whether these regularities are operationally useful, and the answer
separates cleanly. Prediction works: a single historical signal, computable before
re-embedding, cuts retrieval harm substantially relative to both random and
freshness-based scheduling. Combining signals does not. Adding three further signals to
that one leaves the decision statistically unchanged across every budget we evaluate.
Filtering, by contrast, does work: skipping units whose text is unchanged from the
indexed state is safe by construction and avoids up to half of all re-embedding
computations in the most reversion-heavy project.

That contrast is the paper's main empirical lesson. The signals are discriminative and,
with one exception related by construction, unassociated with one another---exactly the
condition under which combining features is standard practice---and combining them still
changes no decision. Signal-level informativeness and decision-level relevance are
different properties, and telling them apart requires an evaluation that scores decisions
rather than signals.

In summary, this work makes four contributions:

\begin{itemize}[leftmargin=*]
    \item \textbf{An empirical characterization of how code changes propagate into retrieval impact.}
    We study the relationship among syntactic change type, embedding displacement, and
    retrieval outcomes across 10 repositories, 8,936 real updates, three programming
    languages, and three embedding models, establishing both the predictive structure
    of this propagation and the recurring counterexamples that bound it.

    \item \textbf{CIPBench, an evaluation that separates informative signals from useful ones.}
    Existing practice validates a maintenance signal by showing it correlates with the
    outcome. CIPBench instead scores the \emph{decision} a signal produces, under a
    strict pre-embedding information boundary, across two tracks and five reference
    baselines. That distinction is what makes the saturation in Section~\ref{sec:trackb}
    visible: the signals that pass a correlational test there still change no outcome.
    The benchmark ships with an open toolchain and human-validated change labels
    (Cohen's $\kappa=0.786$; classifier accuracy $=88.7\%$).

    \item \textbf{A negative result on multi-signal scheduling, with its mechanism.}
    On the full 20-point Track~B grid a weighted combination of four pre-embedding
    signals is statistically equivalent to history alone, and a 130-point sweep spanning
    budgets from 5\% to 100\% finds no budget at which that changes. The combination is
    invariant to the substantive weight and unimproved by the orthogonal exposure
    signal. Cross-validation discards retrieval exposure entirely in eight of the ten
    repositories, so the combination reduces in practice to the single signal it was
    built to improve on.

    \item \textbf{A measured, deployable filter.}
    Track~A shows that hash-identity skipping is safe by construction in all 10
    repositories with savings up to 49.2\%. A Python docstring rule applies to only
    eight units across the benchmark and misses one, too few to validate; we give a
    label-free probing guard for deploying it but cannot exercise that guard on this
    data. Both rules are detectable from the changed text alone, so the filter stage
    costs a hash comparison and a tree-sitter parse and consumes no GPU.
\end{itemize}
\section{Related Work}
\label{sec:related}

Three bodies of work bear on the re-embedding decision: code-change analysis, which
characterizes what changed and what it predicts; code retrieval and repository-level
RAG, which evaluate retrieval quality over a repository snapshot; and dynamic index
maintenance, which incorporates updated vectors into an index efficiently. Each stops
one step short of the decision this paper studies, and for a different reason.

\subsection{Code Change Analysis and Impact Prediction}

Fine-grained code-change analysis has long sought to characterize source-code modifications beyond line-level textual differences. ChangeDistiller extracts fine-grained source-code changes and organizes them into a detailed taxonomy~\cite{fluri2007changedistilling}; Fluri and Gall showed that textual edits do not necessarily correspond to substantive program changes~\cite{fluri2006change}; and Kawrykow and Robillard systematically studied non-essential changes in version histories~\cite{kawrykow2011nonessential}. GumTree later provided a widely used language-independent AST differencing technique~\cite{falleri2014gumtree}, while RefactoringMiner and its extensions identify structured refactorings across successive program versions~\cite{tsantalis2018refactoringminer,tsantalis2022refactoringminer2}. These approaches provide increasingly precise descriptions of what changes between software revisions.

Code changes have also been used to predict downstream maintenance outcomes. Change-impact analysis relies on program dependencies, slicing, and semantic relationships to identify entities that may be affected by a modification~\cite{horwitz1988slicing,bohner1996impact,ren2006semantic}. Fine-grained change features have been studied for defect prediction~\cite{giger2011finegrained} and just-in-time quality assurance~\cite{kamei2013jit}. Historical evolution provides another useful signal: G\^irba \etal's Yesterday's Weather observation shows that entities that changed recently are more likely to continue changing~\cite{girba2004yesterday}.

Predictive test selection is the closest structural analogue to our setting. It estimates which tests are likely to be relevant to a given code change in order to reduce execution cost~\cite{machalica2019predictive,spieker2017reinforcement}, and regression-test selection similarly balances safety against the cost of rerunning a suite~\cite{rothermel1997safe}. Both share the shape of the problem studied here: a change arrives, a budget constrains how much downstream work can be done, and a signal derived from the change must decide what to spend it on. Section~\ref{sec:discussion} returns to this analogy, because the failure mode we identify is one that any budget-constrained selection mechanism can exhibit.

Recent work has extended code-change analysis to large language models. Fan \etal~\cite{fan2025codechange} evaluate LLMs on several code-change-related tasks, including code review generation, commit-message generation, and just-in-time comment updating. Their study examines how effectively LLMs reason about differences between program versions. Our study considers a different downstream consequence of change: whether a modification alters the code representation and, ultimately, the retrieval results produced from that representation.

Prior change-impact studies therefore provide both the change representations and the predictive perspective used in our work, but their outcomes are typically defects, tests, refactorings, or affected program entities. Neural retrieval introduces a different form of impact. Its behavior depends on continuous similarity scores and competition among ranked candidates, so a code unit may move in embedding space without changing any top-$K$ result, while a comparatively small movement may affect retrieval near a ranking boundary. We study this propagation from syntactic change to embedding displacement and retrieval impact.

\subsection{Code Retrieval and Repository-Level RAG}

A large body of work has studied learned representations of source code. CodeBERT~\cite{feng2020codebert}, GraphCodeBERT~\cite{guo2021graphcodebert}, CodeT5~\cite{wang2021codet5}, UniXcoder~\cite{guo2022unixcoder}, and CodeSage~\cite{zhang2024codesage} incorporate programming-language structure, data flow, identifiers, or large-scale code pretraining into code representations. General-purpose encoders such as Sentence-BERT~\cite{reimers2019sentencebert} and MiniLM~\cite{wang2020minilm} are also widely used as transferable embedding backbones. In this work the embedding model is not itself the optimization target; we study how the representation produced by such a model behaves as the underlying source code evolves.

Code-search benchmarks evaluate whether learned representations support effective retrieval. CodeSearchNet~\cite{husain2019codesearchnet} and CoSQA~\cite{huang2021cosqa} provide code--query pairs for semantic code search and question answering. More recent benchmarks and systems move toward repository-level context. RepoBench~\cite{liu2024repobench} and CrossCodeEval~\cite{ding2023crosscodeeval} evaluate repository-level or cross-file code completion, while RepoCoder~\cite{zhang2023repocoder} combines iterative retrieval with code generation. CodeRAG-Bench~\cite{wang2025coderagbench} evaluates the contribution of retrieval to code generation, and general RAG evaluation frameworks such as RAGAS~\cite{es2024ragas} and ARES~\cite{saadfalcon2024ares} assess retrieval and generation quality more broadly. Repository-level RAG research has also investigated how retrieval itself can be improved; Tan \etal~\cite{tan2026procc}, for example, combine multiple retrieval perspectives and adaptively select useful context for code completion. These studies primarily evaluate retrieval over a given repository snapshot.

Recent work has begun to study retrieval under evolving corpora explicitly. 
Zhang \etal~\cite{zhang2025dynamiccorpora} evaluate generative retrieval as new documents are added to a dynamic corpus, while GradNormIR~\cite{ko2025gradnormir} detects corpus-level distribution shift to determine when a dense retriever should be updated. 
FlowRAG~\cite{zhang2026flowrag} instead continually adapts the retriever under non-stationary corpora. 
These works address retrieval adaptation at the corpus or retriever level, rather than deciding whether the representation of an individual changed code unit should be refreshed.

Our setting differs in both granularity and timing. Once an indexed code unit changes, the stored representation may no longer correspond to its current content. Existing code-search and code-RAG benchmarks generally measure whether relevant code can be retrieved from a snapshot; they do not evaluate whether an individual representation should be refreshed after the underlying code changes. Drift-oriented retrieval work such as DeDrift~\cite{baranchuk2023dedrift} studies how search can remain robust \emph{after} content drift has occurred; we instead examine whether the retrieval consequence of an observed change can be anticipated \emph{before} the updated embedding is computed.

\subsection{Dynamic Index Maintenance and Refresh Scheduling}

Dynamic vector indexes support efficient search while the indexed collection changes over time. DiskANN~\cite{subramanya2019diskann} and FreshDiskANN~\cite{singh2021freshdiskann} study graph-based approximate nearest-neighbor search under large-scale and streaming workloads. SPFresh~\cite{xu2023spfresh} supports incremental in-place updates for billion-scale vector search, while Ada-IVF~\cite{mohoney2024adaivf} and Quake~\cite{mohoney2025quake} adapt index organization as vectors are inserted, deleted, or updated. These systems address the structural cost of maintaining a dynamic ANN index.

\emph{A key assumption in these systems is that the updated vector is already available.} Their concern is how that vector should be incorporated into the index efficiently. In code-RAG maintenance, however, obtaining the updated vector may itself be expensive. After observing a source-code change, a system must first decide whether recomputing the corresponding representation is worthwhile. A re-embedding policy can therefore precede dynamic-index maintenance: selected code changes produce new vectors, which are then processed by SPFresh-, Quake-, or graph-based index-maintenance mechanisms.

A complementary line of work reduces representation-maintenance costs under model or distribution drift. 
AURORA~\cite{saravanan2026aurora} adapts retrieval indices under distribution shift through lightweight continual adaptation and uncertainty-aware routing. 
Drift-Adapter~\cite{vejendla2025driftadapter} addresses embedding-model upgrades by mapping queries from a new embedding space into a legacy space, allowing an existing ANN index to remain in service without immediate corpus-wide re-embedding. 
Monath \etal~\cite{monath2024stale} address stale cached embeddings during dense-retriever training with a learned corrector network that reduces repeated re-encoding. 
These approaches address staleness caused by distribution shift, model upgrades, or training-time encoder updates, whereas our setting concerns staleness caused by source-code content changes.

Similar cost--freshness trade-offs appear in other systems domains. Web-crawling policies determine when pages should be revisited in order to maintain freshness under limited bandwidth~\cite{cho2000synchronizing}. Database systems maintain materialized views and cached state after base-data changes~\cite{blakeley1986materialized,franklin1990cache}, while web caching systems study invalidation and consistency for dynamic content~\cite{challenger1999web}. Adaptive cache-replacement methods such as ARC, LeCaR, and LRB further show that inexpensive historical information can support resource-allocation decisions~\cite{megiddo2003arc,vietri2018lecar,song2020lrb}.

The information available in code-RAG maintenance differs from these settings. A web crawler may not know whether a page has changed until it revisits the page, whereas a version-controlled repository exposes the change itself directly. The uncertainty is instead whether that observed change has a meaningful retrieval consequence. Cache invalidation and materialized-view maintenance also commonly rely on discrete dependency relationships, while neural retrieval depends on continuous similarity and ranking geometry. The resulting maintenance decision is therefore not simply whether content changed, but whether the observed change warrants recomputing its representation.

Table~\ref{tab:positioning} places CIPBench against these settings. The decision it formalizes arises only under the conjunction of three conditions that no prior setting combines: code changes continuously, retrieval quality is the quantity at stake, and the new vector does not yet exist.

\begin{table*}[t]
\caption{Comparison with adjacent research settings.}
\label{tab:positioning}
\centering
\scriptsize
\begin{tabularx}{\textwidth}{
    p{0.19\textwidth}
    p{0.17\textwidth}
    p{0.22\textwidth}
    Y}
\toprule
\textbf{Research setting}
&
\textbf{Corpus state}
&
\textbf{Information available}
&
\textbf{Primary question}
\\
\midrule

Code-change analysis
&
Evolving source code
&
Observed code changes and program structure
&
What changed, and which program artifacts or maintenance outcomes may be affected?
\\

Code retrieval and repository-level RAG
&
Typically a fixed repository snapshot
&
Current code representations and retrieval index
&
Which code is relevant to a query, and how can retrieval or generation quality be improved?
\\

Dynamic vector-index maintenance
&
Evolving vector collection
&
Updated vectors are available
&
How should inserted, deleted, or updated vectors be incorporated into the index efficiently?
\\

Refresh and cache scheduling
&
Evolving data or cached state
&
Freshness or change information varies by setting
&
Which state should be refreshed or retained under resource constraints?
\\

\textbf{CIPBench setting}
&
\textbf{Evolving source code}
&
\textbf{Observed code change and existing index, but no updated embedding}
&
\textbf{Should a changed unit be re-embedded, and which units should be prioritized under a limited budget?}
\\

\bottomrule
\end{tabularx}
\end{table*}
\section{Empirical Analysis}
\label{sec:empirical}

We investigate how code changes propagate through the three layers introduced in Section~\ref{sec:intro}: \emph{syntactic substantiality} $\rightarrow$ \emph{embedding displacement} $\rightarrow$ \emph{retrieval impact}. The central question is whether information available when a code change occurs can anticipate its downstream retrieval consequence before the updated embedding is computed. We organize the analysis around four research questions:

\begin{itemize}[leftmargin=*]
    \item \textbf{RQ1---Change and representation:} How does the syntactic nature of a code change relate to embedding displacement?
    \item \textbf{RQ2---Representation and retrieval:} When does embedding displacement translate into retrieval impact?
    \item \textbf{RQ3---Context and predictability:} Which temporal, retrieval-exposure, and repository-level signals provide information about retrieval impact before re-embedding?
    \item \textbf{RQ4---Failure boundaries:} Under what conditions do the three layers fail to align?
\end{itemize}

The remainder of this section maps each research question to a subsection and a numbered finding---RQ1 to Finding~1, RQ2 to Finding~2, RQ3 to Findings~3--5, RQ4 to Finding~6---each supported by a table or figure.

\subsection{Study Setup}\label{subsec:empirical-setup}

The study covers 10 open-source repositories in Python, Go, and JavaScript, comprising 2,366 function-level units and 8,936 real updates organized as issue-driven evolution streams, with 74--983 ground-truth queries per repository. Section~\ref{sec:cipbench} documents how these streams are constructed and released. This subsection defines the three quantities measured for every update, which are used throughout the paper.

We record three quantities, which are not all defined at the same granularity. Change type is a property of an \emph{update}: each update in a unit's history carries a syntactic category. Displacement and retrieval impact are properties of a \emph{unit}: one $\delta$ and one $I$ per unit, measuring net movement and net retrieval consequence across that unit's whole observed history. The decision a maintenance policy makes is likewise per unit (Section~\ref{sec:cipbench}), so where an analysis relates change type to the other two it attributes to each unit the most severe class in its history.

\paragraph{Change type.}
A tree-sitter-based classifier assigns six fine-grained change levels. For readability these are grouped into three analysis tiers: \emph{non-substantive}, \emph{implementation-body}, and \emph{function-signature} changes, where signature changes include modifications to parameters, return types, and type annotations. A lexical fallback is used only when syntactic parsing is unavailable, and accounts for at most 2.8\% of updates in any repository.

\paragraph{Embedding displacement.}
Let $E(\cdot)$ be the encoder mapping source text to a vector. Displacement is measured per \emph{unit} rather than per update: writing $x_{\mathrm{first}}$ for a unit's source text before its earliest observed update and $x_{\mathrm{last}}$ for its text after the latest, we define
\[
\delta = 1-\cos\!\left(E(x_{\mathrm{first}}),\,E(x_{\mathrm{last}})\right),
\]
the cosine distance between where the unit's representation started and where it ended. This is $0$ when the two coincide and grows toward $2$ as they become dissimilar. Measuring net movement rather than the size of individual steps is what makes the reversion cases of Section~\ref{subsec:funnel} visible: a unit edited and then restored has $\delta=0$ exactly, however large the intermediate steps were. The per-update step displacements are retained separately and are used only where movement must be decomposed across a sequence, as in the cumulative-drift case of Section~\ref{subsec:counterexamples}.\footnote{The encoders used here L2-normalize their output vectors, and retrieval scores in the index are themselves cosine similarities; measuring displacement as a cosine distance therefore keeps the representation-level quantity consistent with the retrieval-level quantity it is intended to explain. The form $1-\cos$ also anchors ``no change'' at exactly $0$, which matters for the net-zero reversion cases in Section~\ref{subsec:counterexamples}.} The primary experiments use MiniLM-L6-v2~\cite{wang2020minilm};MPNet-base-v2~\cite{song2020mpnet} and BGE-base-en-v1.5~\cite{xiao2023cpack} are used for cross-model robustness checks.
% The primary experiments use MiniLM-L6-v2~\cite{xxx}; MPNet-base-v2~\cite{xxx} and BGE-base-en-v1.5~\cite{xxx} are used for cross-model robustness checks.

\paragraph{Retrieval impact.}
Impact compares top-$K$ retrieval before and after replacing the changed unit's vector, with $K=10$. Let $G(q)$ denote the ground-truth units of query $q$ and
\[
\mathrm{Hit}_K(q;\mathcal{I}) \;=\; G(q)\,\cap\,\mathrm{Top}_K(q;\mathcal{I})
\]
the ground-truth units that $\mathcal{I}$ actually retrieves for $q$. A unit is impacted, $I>0$, if some ground-truth query satisfies
\[
\mathrm{Hit}_K(q;\mathcal{I}_{\mathrm{old}}) \;\neq\; \mathrm{Hit}_K(q;\mathcal{I}_{\mathrm{new}}),
\]
and $I=0$ otherwise. The predicate therefore records changes in \emph{which} ground-truth units are retrieved, not reorderings among them: a unit that moves from rank~3 to rank~7 while remaining in the top-$K$ does not count as impacted, because no user-visible result set has changed. Section~\ref{sec:threats} discusses the sensitivity of this choice.

Computing retrieval impact requires traversing each repository's query set for every evaluated update, roughly $15$ million top-$K$ retrieval calculations across the three encoders.

The change labels were independently validated on 300 updates, stratified across the 10 repositories. Inter-rater agreement is Cohen's $\kappa=0.786$ (95\% CI [0.729, 0.840]), and the classifier agrees with the adjudicated final labels on 266/300 cases (88.7\%). Section~\ref{sec:threats} discusses the disagreement structure in detail.

Statistical comparisons are performed within repositories wherever possible: Kruskal--Wallis with pairwise Mann--Whitney tests at Bonferroni-adjusted $\alpha'=0.017$ for the displacement classes, and Fisher's exact test for sparse contingency tables. Continuous signals are split at their within-repository medians, except $k$-occ, for which zero exposure and $k$-occ$\ge3$ form operational low- and high-exposure groups; these cutoffs support descriptive comparison and are not deployment thresholds.

\subsection{RQ1: How Does Change Type Relate to Embedding Displacement?}\label{subsec:funnel}

We first examine the link between syntactic change and movement in representation space. If different change types induce systematically different displacement, syntax provides information about representation change without requiring the updated embedding itself.

Each unit contributes one displacement, attributed to the most severe change class observed anywhere in its history; Table~\ref{tab:delta-by-change} reports the median over the units in each class.
The ratio compares the larger of the median signature- or implementation-change displacement with the median non-substantive displacement. Categories with $n<5$ are excluded from this ratio to avoid unstable denominators.

\begin{table*}[t]
\caption{Median embedding displacement across three change classes in 10 repositories.}
\label{tab:delta-by-change}
\centering
\small
\resizebox{\linewidth}{!}{
\begin{threeparttable}
\begin{tabular}{llcccc}
\toprule
\textbf{Repository} & \textbf{Lang.} &
\textbf{Non-substantive} $\mathrm{med}(\delta)$ &
\textbf{Implementation} $\mathrm{med}(\delta)$ &
\textbf{Signature} $\mathrm{med}(\delta)$ &
\textbf{Ratio}\\
\midrule
flask & Py & 0.019 ($n=4$)\tnote{*} & 0.005 & 0.061 & --\tnote{$\ddagger$}\\
aiohttp & Py & 0.026 ($n=2$) & 0.027 & 0.000\tnote{$\dagger$} & --\\
pylint & Py & 0.000 ($n=9$) & 0.020 & 0.014 & --\\
requests & Py & 0.002 ($n=1$) & 0.006 & 0.049 & --\tnote{$\ddagger$}\\
click & Py & 0.026 ($n=3$)\tnote{*} & 0.040 & 0.031 & --\\
cobra & Go & 0.002 & 0.044 & 0.144 & $85\times$\\
echo & Go & $\approx0$ ($n=1$) & 0.028 & 0.137 & --\\
gin & Go & -- & 0.051 & 0.167 & --\\
lodash & JS & 0.015 & 0.033 & 0.078 & $5\times$\\
express & JS & 0.006 ($n=1$) & 0.035 & 0.156 & --\\
\bottomrule
\end{tabular}
\begin{tablenotes}\footnotesize
\item[*] Two of four flask units and two of three click units have $I>0$; these cases are revisited in Section~\ref{subsec:counterexamples}.
\item[$\dagger$] Of 43 aiohttp signature-change units, 28 have $\delta=0$ exactly.
\item[$\ddagger$] Ratio withheld: the non-substantive category has $n<5$, too few for a stable denominator.
\end{tablenotes}
\end{threeparttable}
}
\end{table*}

Across repositories, signature changes generally induce larger displacement than implementation-body changes, while implementation changes tend to exceed non-substantive edits. The separation can be large: in cobra, the median displacement for signature changes is $85\times$ the non-substantive median. A Kruskal--Wallis test confirms a strong difference across the three classes, and all three pairwise comparisons survive multiplicity correction with rank-biserial effect sizes between 0.39 and 0.68; Appendix~\ref{app:robustness} gives the test statistics. Each unit enters the comparison once, since displacement is a unit-level quantity; the dependence that remains is in the class label rather than the observation count, because a unit whose history spans several classes is assigned only its most severe.

One possible explanation is that signature edits disproportionately modify identifier-bearing tokens---function names, parameter names, and return/type annotations---whereas implementation changes are distributed over a larger token sequence. We treat this as a hypothesis rather than a demonstrated mechanism; direct probing of encoder sensitivity is outside the scope of this study.

The aiohttp results illustrate an important boundary. Of 43 signature-change units, 28 (65\%) have exactly zero net displacement. Inspection reveals repeated $A\rightarrow B\rightarrow A$ patterns in which type annotations are changed and later restored. These reverts are retained in the stream as real repository events; a sensitivity analysis excluding exact text-level reverts is provided in the artifact. Syntactic substantiality therefore indicates that a meaningful edit occurred, but does not guarantee that the final representation differs from its earlier state.

\textbf{Finding 1.}
\emph{Change type is a usable proxy for displacement but not a substitute for it. The class separation is large enough to act on (up to $85\times$ in median displacement), yet 65\% of aiohttp's signature changes move the representation not at all---so syntax identifies edits that \emph{could} matter, not edits that \emph{did}.}

\subsection{RQ2: When Does Embedding Displacement Affect Retrieval?}\label{subsec:retrieval-impact}

The first link in the funnel is strong, but representation movement is useful for maintenance only if it translates into retrieval change. We therefore examine displacement together with the unit's position in the existing retrieval space.

For a unit $u$, we define $k$-occ as the number of ground-truth queries for which $u$ appears among the current top-$K$ results, with $K=10$. Unlike displacement, $k$-occ can be computed entirely from the old index. It captures the unit's observed retrieval exposure: a unit with $k$-occ$=0$ has not appeared in the top-$K$ results of any ground-truth query.

Table~\ref{tab:position-impact} reports retrieval-impact rates after jointly stratifying units by displacement and $k$-occ.

\begin{table}[htbp]
\caption{Joint discrimination of retrieval impact by displacement and retrieval position.}
\label{tab:position-impact}
\centering
\small
\begin{threeparttable}
\begin{tabular}{lcccc}
\toprule
\textbf{Repository} &
$k$-occ$\ge3$ &
$k$-occ$=0$ &
\textbf{High--high} &
\textbf{Low--low}\\
&
$P(I>0)$ &
$P(I>0)$ &
$P(I>0)$ &
$P(I>0)$\\
\midrule
flask & 16.2\% & 12.5\% & 22.6\% & 0\%\\
aiohttp & 13.1\% & 0.0\% & 25.0\% & 0\%\\
pylint & 25.9\% & 0.0\% & 36.6\% & 0\%\\
requests & 23.3\% & N/A\tnote{$\dagger$} & 40.9\% & N/A\tnote{$\dagger$}\\
click & 34.8\% & N/A\tnote{$\dagger$} & 50.0\% & N/A\tnote{$\dagger$}\\
cobra & 39.5\% & 13.0\% & 48.8\% & 0\%\\
echo & 32.5\% & 17.3\% & 49.3\% & 0\%\\
gin & 39.3\% & 16.7\% & 53.3\% & 0\%\\
lodash & 21.0\% & 6.8\% & 26.7\% & 11.8\%\\
express & 34.3\% & N/A\tnote{$\dagger$} & 55.6\% & N/A\tnote{$\dagger$}\\
\bottomrule
\end{tabular}
\begin{tablenotes}\footnotesize
\item[$\dagger$] No unit in requests, click, or express has $k$-occ$=0$; their minimum values are 7, 5, and 1, respectively. Empty cells should not be interpreted as $P(I)=0$.
\end{tablenotes}
\end{threeparttable}
\end{table}

``High--high'' denotes displacement above the repository median together with $k$-occ$\ge3$; ``low--low'' denotes displacement below the median together with $k$-occ$=0$. The $k$-occ$\ge3$ cutoff is used here as an operational high-exposure group rather than as a universal boundary.

The two dimensions separate impacted units substantially better when considered together. Across repositories, the high--high region has $P(I>0)$ between 22.6\% and 55.6\%. In six repositories with sufficient low-exposure support---aiohttp, pylint, flask, cobra, echo, and gin---the low--low region contains no impacted unit; lodash is the exception at 11.8\%. Per-repository Fisher tests are significant at $\alpha=0.05$ in 9/10 repositories and remain strong under the conservative Bonferroni threshold of $\alpha'=0.005$. The corresponding odds ratios range from 3.5 to 23.2.

The evidence therefore does not support a monotonic ``larger displacement implies retrieval impact'' interpretation. A unit can move substantially without altering a relevant ranking if it has little retrieval exposure, whereas comparatively small movements may matter when a unit already competes near a top-$K$ boundary.

Two mechanisms are consistent with this pattern. First, rankings near the top-$K$ boundary can be sensitive to small geometric changes. Second, highly exposed units may lie in competitive neighborhoods where several candidates have similar scores. The present experiment establishes the positional association; direct measurement of boundary geometry is left for future work.

\textbf{Finding 2.}
\emph{Retrieval position is a distinct signal, not a refinement of displacement. Even an oracle with perfect knowledge of how far an embedding will move cannot say where that movement matters: the two dimensions jointly reach $P(I>0)=22.6$--$55.6\%$ where either alone is uninformative.}

\subsection{RQ3: Which Pre-Embedding Signals Are Informative?}\label{subsec:context}

RQ1 and RQ2 identify two sources of pre-embedding information: the current syntactic change and the unit's existing retrieval exposure. We now examine two additional sources---change history and repository context---and then compare the four inexpensive signals directly.

\subsubsection{Temporal Persistence}

Prior software-evolution work has reported persistence in change behavior. In particular, G\^irba \etal's Yesterday's Weather observation suggests that entities undergoing substantial change in the recent past are likely to remain change-prone in subsequent versions~\cite{girba2004yesterday}. Motivated by this observation, we test a related form of persistence for substantive updates.

For every unit with multiple updates, we compute the probability that later updates are substantive conditional on whether the first observed update is substantive. Table~\ref{tab:yesterday} reports the results.

\begin{table*}[t]
\caption{Temporal persistence of substantive updates across 10 repositories.}
\label{tab:yesterday}
\centering
\small
\resizebox{\linewidth}{!}{
\begin{threeparttable}
\begin{tabular}{llrrrr}
\toprule
\textbf{Repository} & \textbf{Lang.} &
\textbf{Multi-update units} &
\textbf{First subst.$\rightarrow$later subst.} &
\textbf{First non-subst.$\rightarrow$later subst.} &
\textbf{Effect (pp)}\\
\midrule
aiohttp & Py & 52 & 99.3\% & 20.4\% & +78.9\\
click & Py & 16 & 93.9\% & 0.0\% & +93.9\\
pylint & Py & 82 & 91.1\% & 25.2\% & +66.0\\
requests & Py & 11 & 96.1\% & 50.0\% & +46.1\\
flask & Py & 87 & 92.6\% & 55.6\% & +37.1\\
cobra & Go & 173 & 92.1\% & 57.7\% & +34.4\\
echo & Go & 436 & 96.9\% & 69.8\% & +27.1\\
gin & Go & 263 & 94.9\% & 93.5\% & +1.4\tnote{$\ddagger$}\\
lodash & JS & 194 & 97.7\% & 77.5\% & +20.2\\
express & JS & 23 & 97.7\% & 100.0\% & -2.3\tnote{$\ddagger$}\\
\bottomrule
\end{tabular}
\begin{tablenotes}\footnotesize
\item[$\ddagger$] The comparison group contains fewer than five units in gin and express, both of which have very high substantive-update rates; 6/10 per-repository Fisher tests are significant at $\alpha=0.05$.
\end{tablenotes}
\end{threeparttable}
}
\end{table*}

Across repositories, a unit whose first observed update is substantive has a 91.1--99.3\% probability that later updates are also substantive. The contrast with initially non-substantive units varies considerably. aiohttp, click, and pylint show differences of +66 to +94 percentage points, whereas gin and express provide little contrast because nearly all updates in those repositories are substantive.

At first glance the larger effects in several Python repositories might suggest a language effect. Persistence of the exact coarse change class between consecutive updates rules that out (Appendix~\ref{app:persistence}): the language-level averages of 84.5\%, 71.5\%, and 77.7\% for Python, Go, and JavaScript hide larger within-language variation, with aiohttp, click, and pylint at 93--96\% but flask and requests at 65--73\%, overlapping the Go and JavaScript repositories.

One plausible repository-level explanation is gradual type-hint adoption: annotation-heavy libraries may repeatedly refine function signatures, creating long runs of signature-to-signature changes, while repositories with more heterogeneous feature and bug-fix evolution show weaker persistence. The mechanism remains a hypothesis, but the empirical result is clear: recent change history contains useful information, and its strength is repository-dependent rather than language-dependent.

\textbf{Finding 3.}
\emph{Substantive changes exhibit substantial temporal persistence. In repositories with informative comparison groups, the lift associated with an initially substantive update ranges from roughly +20 to +94 percentage points, but the strength of this effect varies more by repository than by language.}
This persistence motivates using history as a prior, but not as a repository-independent rule.

\subsubsection{Repository Temperature}

The repositories also differ in the overall fraction of updates that are substantive (Table~\ref{tab:temperature}); we refer to this update-level fraction as the repository's \emph{temperature}.

\begin{table}[htbp]
\caption{Update-level temperature (substantive fraction of updates) across the 10 repositories. Headroom is the non-substantive fraction available for change-based filtering.}
\label{tab:temperature}
\centering
\small
\begin{tabular}{lrrrr}
\toprule
\textbf{Repository} & \textbf{Lang.} & \textbf{Updates} & \textbf{Temperature} & \textbf{Headroom}\\
\midrule
pylint & Py & 720 & 70.6\% & 29.4\%\\
cobra & Go & 1,079 & 80.5\% & 19.5\%\\
requests & Py & 69 & 87.0\% & 13.0\%\\
click & Py & 86 & 87.2\% & 12.8\%\\
flask & Py & 334 & 87.4\% & 12.6\%\\
aiohttp & Py & 409 & 89.5\% & 10.5\%\\
gin & Go & 1,802 & 95.7\% & 4.3\%\\
lodash & JS & 950 & 95.9\% & 4.1\%\\
echo & Go & 3,351 & 96.0\% & 4.0\%\\
express & JS & 136 & 97.8\% & 2.2\%\\
\bottomrule
\end{tabular}
\end{table}

Temperature changes the opportunity available to any selective maintenance method. In a repository with roughly 70\% substantive updates, close to 30\% of updates remain potential candidates for avoiding recomputation based on change substantiality. When more than 95\% of updates are substantive, the corresponding headroom is below 5\%. The available non-substantive headroom therefore differs by more than an order of magnitude across the observed workloads, from 29.4\% (pylint) to 2.2\% (express).

Two distinct quantities are easy to conflate here. Temperature is computed over \emph{updates} and bounds the headroom for change-based filtering. A separate quantity, the substantive fraction of \emph{retained units}, ranges from 87.0\% (click) to 99.1\% (gin) and instead determines whether the substantive tier exceeds a scheduling budget; Section~\ref{sec:trackb} uses it for that purpose. The two should not be substituted for one another: pylint is 70.6\% by updates but 92.6\% by units.

This distinction is independent of whether a particular maintenance policy is effective: a hot repository simply contains fewer obviously low-substantiality updates to exploit. Temperature is consequently useful for interpreting cross-repository differences in attainable savings.

\textbf{Finding 4.}
\emph{Repositories differ substantially in their evolution ``temperature.'' This workload property directly constrains the headroom available to change-based filtering and helps explain why identical decision rules can have different opportunities across repositories.}

\subsubsection{Discrimination of the Four Pre-Embedding Signals}\label{subsec:signals}

We next compare the four signals available at the refresh point, before any updated
embedding is computed. They are not all of the same granularity: change type is a
property of an individual update, which we aggregate to the unit as described above,
while the other three are unit-level quantities.

\begin{itemize}[leftmargin=*]
    \item \texttt{change\_type}: the syntactic category of the current update, taking one of the three analysis tiers. This is distinct from the binary substantive flag used by the policy of Section~\ref{sec:trackb}, which is a unit-level property rather than a property of the current update, and the two behave differently: in requests they are associated with $n_{sub}$ at $-0.170$ and $+0.475$ respectively;
    \item $k$-occ: the unit's exposure in the existing top-$K$ retrieval results;
    \item $n_{sub}$: the number of substantive updates in the unit's observed history. Because the decision object is the unit (Section~\ref{sec:cipbench}), this is a count over the unit's whole stream rather than a prefix of it; there is no ``current'' update to exclude. It is computable before any embedding, since change type is derived from source text alone; and
    \item \texttt{nn\_sim}: the cosine similarity between the unit's old embedding and its nearest neighbor in the old index.
\end{itemize}

For descriptive comparison, we measure signal discrimination as $\Delta P = P(I>0\mid\mathrm{high}) - P(I>0\mid\mathrm{low})$. Positive values indicate that the high-signal group contains a larger fraction of impacted units. We use $\Delta P>0.10$ to highlight practically noticeable separation in this analysis; it is an operational descriptive cutoff rather than a universal statistical threshold.

\begin{table}[htbp]
\caption{Discriminative power $\Delta P=P(I>0\mid high)-P(I>0\mid low)$ of four pre-embedding signals.}
\label{tab:signal-dp}
\centering
\small
\begin{tabular}{llrrrr}
\toprule
\textbf{Repository} & \textbf{Lang.} &
\textbf{change\_type} &
\textbf{$k$-occ} &
\textbf{$n_{sub}$} &
\textbf{nn\_sim}\\
\midrule
flask$^s$ & Py & -0.083 & +0.114 & +0.086 & +0.033\\
aiohttp & Py & +0.133 & +0.098 & +0.088 & -0.030\\
pylint & Py & +0.227 & +0.164 & +0.191 & +0.007\\
requests$^s$ & Py & -0.280 & +0.197 & +0.093 & -0.023\\
click$^s$ & Py & -0.367 & +0.049 & +0.078 & +0.092\\
cobra & Go & +0.265 & +0.100 & +0.183 & -0.007\\
echo & Go & +0.280 & +0.052 & +0.196 & -0.049\\
gin & Go & +0.131 & +0.081 & +0.037 & -0.005\\
lodash & JS & +0.165 & +0.026 & +0.078 & +0.002\\
express & JS & +0.343 & +0.098 & +0.147 & -0.024\\
\bottomrule
\end{tabular}
\vspace{2pt}
\begin{minipage}{0.96\linewidth}\footnotesize
$^s$ Small-$N$: $n_{nonsub}\le16$. Per-signal sample counts are retained in the benchmark data.
\end{minipage}
\end{table}

The signals separate on stability rather than on strength. \texttt{change\_type} clears the cutoff most often (7/10) and has the largest individual values, but it reverses sign in flask, requests, and click---precisely the three repositories with at most 16 non-substantive examples, where a handful of impacted units can flip the estimated polarity---and its mean is dragged to $+0.081$ as a result. $k$-occ and $n_{sub}$ clear the cutoff less often (3/10 and 4/10) but never reverse sign in any repository, and their means are comparable or higher ($+0.098$ and $+0.118$). \texttt{nn\_sim} clears the cutoff in no repository, reaching at most $+0.092$, and averages to zero.

The ordering by mean is therefore $n_{sub}$, $k$-occ, \texttt{change\_type}, \texttt{nn\_sim}, while the ordering by cutoff count puts \texttt{change\_type} first. The disagreement is the point: a signal can be the strongest where it works and still be the least dependable, and a maintenance policy has to live with the latter.

The signals also differ in redundancy. The mean absolute Kendall association between $k$-occ and the other signals is below 0.15, and a per-repository Kendall-$\tau$ analysis (Appendix~\ref{app:signal-independence}) confirms this at the unit level: the language-level association between $k$-occ and $n_{sub}$ is weak ($\tau=0.13$, $0.11$, and $0.24$ for Python, Go, and JavaScript, respectively), whereas the association between \emph{change type} and $n_{sub}$ is moderate on average ($\tau=0.32$, $0.24$, $0.22$). The language means conceal wide within-language variation: the association reaches $+0.52$ in click and $+0.48$ in flask but is essentially absent in gin ($+0.05$) and negative in requests ($-0.17$). History therefore overlaps substantially with change type in most repositories without subsuming it everywhere, and the exceptions are the repositories in which the two signals would be expected to contribute independently. Nearest-neighbor similarity, in contrast, provides little consistent discrimination.

\textbf{Finding 5.}
\emph{No single inexpensive signal dominates across repositories. Change type is often the strongest individual discriminator but becomes unstable when non-substantive evidence is sparse; $k$-occ is more consistently positive and comparatively independent, while nearest-neighbor similarity contributes little on average.}

The observed sign reversals motivate treating small non-substantive samples conservatively; we do not infer a deployment threshold directly from the largest observed sample of 16.

Repository-level meta-features---size, temperature, annotation density, history length, and impacted-unit fraction---show associations with per-signal strength that we report in Appendix~\ref{app:meta-features}. Because those correlations rest on only 10 repositories and we do not act on them in any policy, we treat them as exploratory context rather than as a finding.

\subsection{RQ4: Where Does the Three-Layer Relationship Break Down?}\label{subsec:counterexamples}

Aggregate relationships can hide precisely the cases that matter most to a maintenance decision. We therefore examine recurring mismatches among syntax, embedding displacement, and retrieval impact. They are not isolated noise: each exposes a distinct limitation of single-signal reasoning. Table~\ref{tab:counterexamples} summarizes the recurring failure types and their maintenance implications. The case studies below come from a targeted inspection of extreme recall-loss events and of units whose syntactic class and retrieval outcome disagree; they are intended as design constraints, not as an exhaustive census of failures.

\begin{table*}[t]
\caption{Recurring counterexamples to the syntax--displacement--impact chain.}
\label{tab:counterexamples}
\centering
\small
\resizebox{\linewidth}{!}{
\begin{tabular}{lp{0.40\textwidth}l}
\toprule
\textbf{Type} & \textbf{Evidence} & \textbf{Maintenance implication}\\
\midrule
Embedding moves, retrieval does not & docstring-only: flask 5 units ($\mathrm{med}\,\delta=0.024$, $P(I){=}0$); aiohttp 5 units ($P(I){=}0$) & docstring filter candidate under a per-repository guard\\
Executable syntax unchanged, retrieval changes & comment-only: flask 2/4, click 2/3 units impacted & comments remain rankable; no comment filter\\
Cross-language variation & Go string-only: cobra 1/5, gin 1/4 units impacted & no universal non-substantive rule\\
Net-zero reversion & aiohttp 28/43 signature units have $\delta{=}0$ ($A\rightarrow B\rightarrow A$) & inexpensive hash-equality skip\\
Cumulative drift & flask query ``limits'': $\delta{=}0.177$ over 4 updates; recall 1.00$\rightarrow$0.00 & per-update gating insufficient; periodic probing\\
Top-$K$ boundary & flask issue \#2935: $\delta{=}0.055/0.019$; recall 1.00$\rightarrow$0.50 & displacement alone insufficient (Finding~2)\\
\bottomrule
\end{tabular}
}
\end{table*}

\subsubsection{Four Mismatches, Four Different Causes}

Table~\ref{tab:counterexamples} gives the evidence for the first four failure types. What it cannot give is why each occurs, and the reasons do not generalize to one another.

Docstring-only edits move the embedding without moving retrieval because the affected units have low exposure: a representation can shift noticeably while leaving the top-$K$ results unchanged if the unit rarely enters a relevant query neighborhood. This reinforces Finding~2 and marks the class as low-risk in the studied Python repositories---on five units per repository, evidence enough for a guarded candidate rule but not a universal one. Comment-only edits produce the opposite mismatch for a model-related reason: the primary encoder represents the full text of a unit, so natural-language terms in a comment move the embedding even when the executable AST is untouched. Under text-oriented encoders, syntactic non-substantiality does not imply retrieval non-substantiality. Go string-literal edits then show that one non-substantive category can behave differently across languages, because error messages such as ``failed to parse config'' overlap directly with query terms and template strings can carry identifiers.

Net-zero reversions are qualitatively unlike the other three. They separate \emph{change occurrence} from \emph{net representational change}, and because the final equality of the code state is detectable without computing an embedding, this is the one failure mode admitting an exact and inexpensive test.

\subsubsection{Large Recall Losses from Cumulative and Positional Effects}

Three flask queries exhibit particularly large recall drops, but through two different mechanisms.

The first is cumulative drift. For the query ``add config and docs for limits,'' associated with \code{flask/wrappers.py::max\_content\_length}, four consecutive updates accumulate a displacement of $\delta=0.1773$, while the largest individual step is 0.1118. Recall first collapses at $t=4$, falling from 1.00 to 0.00.

This exposes a limitation of per-update gating: individually moderate changes can accumulate until their combined displacement becomes consequential, and a rule evaluating each update independently may not detect the pattern before the accumulated drift crosses a retrieval boundary.

The remaining two cases arise through position rather than accumulation. Two units associated with issue \#2935 fail at their first observed update, with displacements of only 0.0546 and 0.0192, yet recall falls from 1.00 to 0.50---the top-$K$ boundary effect suggested in RQ2, where a small movement matters because the unit occupies a sensitive ranking position.

Together, the three cases account for 2.3\% of flask's 128 ground-truth queries. They are rare, but they expose two different limitations of displacement-based reasoning: cumulative movement over time and sensitivity to retrieval position.

\textbf{Finding 6.}
\emph{The mismatches among syntax, displacement, and retrieval impact have recurring structure (Table~\ref{tab:counterexamples}). We observe retrieval-irrelevant embedding movement, text-sensitive non-substantive changes, cross-language variation, net-zero reversions, cumulative drift, and top-$K$ boundary effects. No single layer is therefore sufficient for reliable maintenance decisions.}

These counterexamples provide design constraints rather than ready-made policy rules. They motivate conservative treatment of apparently low-impact changes, an explicit net-zero check, and a separate mechanism for cumulative drift.

\subsection{Summary of Findings}\label{subsec:empirical-summary}

The empirical results support the three-layer funnel as an organizing model, but not as a deterministic chain. Change type predicts how far a representation moves (Finding~1); retrieval exposure determines whether that movement becomes consequential (Finding~2); change history supplies an additional temporal prior (Finding~3); repository temperature bounds the headroom available to any change-based rule (Finding~4); and of the four inexpensive signals three discriminate impacted units while neighbourhood similarity does not, with only one associated pair among the six (Finding~5). The counterexamples then show why no single layer suffices (Finding~6).

Findings~3 and~5 together constitute a prior worth testing: if three of the four signals discriminate impacted units and five of the six pairings are unassociated, combining them should improve a maintenance decision. Section~\ref{sec:cipbench} builds the benchmark needed to test that prior under an explicit pre-embedding information boundary; Section~\ref{sec:trackb} tests it and finds it does not hold, and Section~\ref{sec:tracka} shows which part of the evidence does convert into a deployable rule.
\section{CIPBench: A Benchmark for Re-embedding Decisions}\label{sec:cipbench}

Section~\ref{sec:empirical} shows that code changes differ systematically in their embedding and retrieval consequences, but also that no single observable signal perfectly identifies consequential updates. CIPBench formalizes this problem as a benchmark for deciding \emph{whether and when a changed code unit should be re-embedded}. The benchmark records the eventual representation and retrieval consequences of each update for evaluation, while restricting candidate methods to information available before the updated embedding is computed.

The measurements are defined in Section~\ref{subsec:empirical-setup} and are not restated here; recall that change type is recorded per update while displacement and impact are recorded per unit. This section contributes what is specific to the benchmark: the information boundary that separates deployable methods from oracles, the construction of the evolution streams, the two evaluation tracks and their metrics, and the reference baselines.

\subsection{Task Definition and Information Boundary}\label{subsec:cipbench-protocol}

Consider an evolving repository with a sequence of code updates $u_1,u_2,\ldots,u_T$ distributed over a set of code units. The \emph{decision object is the unit}: a maintenance policy decides whether to recompute a unit's representation, and CIPBench scores that decision. Updates are the evidence stream from which the decision is made---they supply each unit's change history---but they are not themselves scored. Accordingly the benchmark records, for each unit, its cumulative displacement $\delta$ and its retrieval impact $I$, together with the sequence of change types observed across its updates.

\begin{figure}[t]
\centering
\includegraphics[width=0.5\columnwidth]{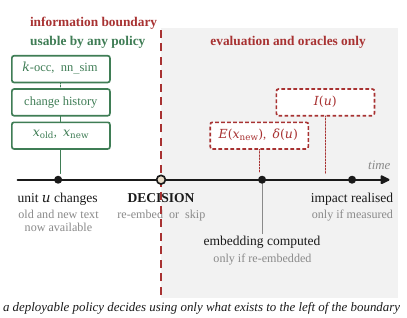}
\caption{The CIPBench information boundary. At a refresh point a unit's accumulated changes and its retrieval context are known, but the decision precedes embedding computation, so $\delta$ and $I$ do not yet exist when a deployable method must choose; they are available only to evaluation and to oracle baselines.}
\label{fig:boundary}
\end{figure}

The central constraint is temporal and operational. At the time a maintenance decision is made, a method may use
\begin{itemize}[leftmargin=*]
    \item the old and updated source text;
    \item repository metadata and the unit's accumulated change history up to the refresh point; and
    \item the state of the existing retrieval index.
\end{itemize}
It may \emph{not} use the updated embedding, the true displacement $\delta(u)$, or the true retrieval impact $I(u)$. These quantities become available only after performing the computation that the maintenance decision is intended to avoid, and are therefore reserved for evaluation or oracle baselines. Figure~\ref{fig:boundary} places each quantity at the point in time at which it comes into existence.

This information boundary distinguishes the prediction problem from ordinary post-update analysis, and it is representational rather than temporal. CIPBench models \emph{periodic} index maintenance: at a refresh point, a set of units has accumulated changes since the last refresh, and the policy decides for each whether to recompute its vector. The decision point is therefore after that unit's accumulated evolution is known and before any embedding is computed from it. What the boundary excludes is everything derived from the new representation---$E(x_{\mathrm{new}})$, $\delta$, and $I$---not the earlier part of the unit's own change history. Repository evolution beyond the refresh point remains unavailable, as it must: no feature may be computed from updates the policy has not yet seen.

\subsection{Dataset Construction}\label{subsec:cipbench-data}

\subsubsection{Issue-Driven Evolution Streams}

CIPBench is constructed from issue-driven repository evolution. Each issue provides an externally meaningful evolution event: its title and body form a natural-language retrieval query, while the code units associated with the issue form the corresponding ground-truth target set.

This construction serves two purposes. First, it preserves a connection between repository evolution and an explicit retrieval intent. Second, it reduces the number of updates for which no issue-level semantic context is available. The resulting stream is not intended to reproduce an unconstrained production commit stream, and this distinction is treated as an external-validity boundary in Section~\ref{sec:threats}.

The benchmark contains 10 repositories in Python, Go, and JavaScript. Table~\ref{tab:cipbench-repos} summarizes the retained function-level units and updates. Repositories carry between 74 and 983 ground-truth queries. That range matters for reading the results: $\mathrm{Hurt@B}$ is a fraction of a repository's own query set, so a given harm value rests on a denominator that differs by more than an order of magnitude across the benchmark, and a small value in a query-poor repository can correspond to a single affected query.

\begin{table}[htbp]
\caption{CIPBench repositories and evolution-stream size.}
\label{tab:cipbench-repos}
\centering
\small
\begin{tabular}{llrr}
\toprule
\textbf{Repository} & \textbf{Lang.} & \textbf{Units} & \textbf{Updates}\\
\midrule
flask & Py & 131 & 334\\
aiohttp & Py & 65 & 409\\
pylint & Py & 162 & 720\\
requests & Py & 43 & 69\\
click & Py & 23 & 86\\
\midrule
cobra & Go & 259 & 1,079\\
echo & Go & 678 & 3,351\\
gin & Go & 434 & 1,802\\
\midrule
lodash & JS & 535 & 950\\
express & JS & 36 & 136\\
\midrule
\textbf{Total} & & \textbf{2,366} & \textbf{8,936}\\
\bottomrule
\end{tabular}
\end{table}

Repositories were selected to cover active projects over multi-year histories, at least 100 issue-level events before benchmark filtering, a broad range of retained repository sizes, and multiple projects within each language. The resulting benchmark contains between 23 and 678 retained units per repository.

\subsubsection{Construction Pipeline}

Benchmark construction proceeds in three stages. First, tree-sitter extracts function- and method-level code units and aligns them across repository evolution. Second, real updates to the retained units are ordered along the issue-driven evolution stream. Third, issue titles and bodies are converted into retrieval queries, with the associated code units serving as retrieval ground truth.

Each retained unit is then annotated with the measurements of Section~\ref{subsec:empirical-setup}, and each update with its change type. Two properties of that annotation are specific to the benchmark release. Displacement is recorded per unit, between its first and last observed states, and the release additionally stores the per-update step displacements so that movement can be decomposed across a sequence where required; the cumulative-drift case of Section~\ref{subsec:counterexamples} uses both, and states which. And because impact labels are sensitive to $K$, the released pipeline supports regeneration at other values of $K$ and with graded relevance measures; $K=10$ is the default used throughout this paper. We denote the set of impacted units by
\[
H=\{\,\text{units with } I>0\,\}.
\]
Computing these labels requires traversing the repository query set for each evaluated unit, approximately 15 million top-$K$ retrieval calculations over the complete benchmark.

\subsubsection{Label Quality and Reproducibility}

The change classifier was evaluated with 24 cross-language smoke tests and a manually validated sample of 300 updates, stratified as 30 updates per repository, which two annotators labeled independently before disagreement adjudication. Section~\ref{subsec:empirical-setup} reports the resulting agreement figures, and Section~\ref{sec:threats} reports the disagreement structure and repository-level variation.

The standard impact labels are computed with the primary encoder MiniLM-L6-v2. Because impacted sets transfer only partially across encoders (Section~\ref{sec:threats}), CIPBench should be read as parameterized by encoder: labels for a target encoder must be regenerated with the released pipeline rather than transferred.

The benchmark artifact includes the data representation, loaders, evaluation scripts, and the measurement pipeline, so that future maintenance policies can be evaluated under the same data construction and information constraints.

\subsection{Evaluation Protocol}

CIPBench evaluates two maintenance tasks. Track~A asks whether a changed unit can be skipped altogether; Track~B asks how the remainder should be prioritized when only a limited fraction can be re-embedded. We present them in that order, which is the order in which a deployed system encounters them and the order of the evaluations in Sections~\ref{sec:tracka} and~\ref{sec:trackb}. Track~A's metrics speak directly to computational savings; Track~B carries the reference baselines and the leaderboard below, and is where the benchmark's discriminative power is established.

\subsubsection{Track A: Impact-Prediction Filtering}

Track~A evaluates selective re-embedding as a binary decision. Let $U$ denote the set of evaluated units and $S\subseteq U$ the units a policy selects for re-embedding. We report two complementary metrics,
\[
\mathrm{MissRate}
=
\frac{|H\setminus S|}{|H|},
\qquad
\mathrm{Savings}
=
1-\frac{|S|}{|U|}.
\]

Miss rate measures the fraction of truly consequential units that the policy fails to re-embed; savings measures the fraction of embedding computations avoided. A useful filtering policy trades one against the other.

We report this pair rather than conventional classification accuracy because retrieval-impact labels are imbalanced and the operational costs of the two error types are asymmetric. Skipping an unaffected unit saves computation; skipping an impacted unit can degrade retrieval quality.

\subsubsection{Track B: Budget-Constrained Scheduling}

Track~B considers settings in which only a fraction of units can be re-embedded. For a budget $B\in\{10,20,\ldots,100\}\%$, a policy ranks the available units and selects the highest-ranked $B\%$ for re-embedding. As in Track~A the budget is a fraction of units, not of updates: pylint's 162 units at $B{=}30\%$ give 48 re-embeddings.

Let $S_B \subseteq U$ denote the set of units selected under budget $B$, and let $\mathcal{I}^{\mathrm{full}}$ and $\mathcal{I}^{S_B}$ denote the indexes in which, respectively, every evaluated unit and only the units in $S_B$ have been re-embedded. Writing
\[
\mathrm{drop}(q) \;=\;
\mathrm{Recall@K}\!\left(q;\mathcal{I}^{\mathrm{full}}\right)
- \mathrm{Recall@K}\!\left(q;\mathcal{I}^{S_B}\right),
\]
with $\mathrm{Recall@K}(q;\mathcal{I}) = |G(q)\cap\mathrm{Top}_K(q;\mathcal{I})|\,/\,|G(q)|$, the primary metric is the fraction of queries that lose a material amount of recall,
\[
\mathrm{Hurt@B}(S_B) \;=\;
\frac{1}{|Q|}\,\bigl|\{\,q \in Q : \mathrm{drop}(q) \ge 0.1 \,\}\bigr| .
\]
Lower values indicate better preservation of retrieval quality. We count queries rather than averaging the drop because the two behave differently when a query has several ground-truth units: averaging lets many small partial losses offset a query that loses its only retrieved match, whereas the operational concern is how many queries are visibly degraded. The $0.1$ threshold excludes losses smaller than one ground-truth unit in a ten-unit target set. CIPBench additionally records the number of affected queries and the maximum per-query recall loss, allowing aggregate harm to be considered together with worst-case behavior.

The protocol is defined over the complete budget range. For compact presentation, the main tables report $B=30\%$ and $B=50\%$ as representative constrained operating points.

\subsection{Reference Baselines}

CIPBench includes five reference policies spanning freshness-based, retrieval-aware, history-based, and oracle information.

\begin{enumerate}[leftmargin=*]

    \item \textbf{Random.}
    Updates are randomly ordered. Results are averaged over five random seeds.

    \item \textbf{Recency.}
    Updates are prioritized by recency, representing a freshness-oriented policy that does not inspect the semantic nature of the change.

    \item \textbf{$k$-occ.}
    Updates are prioritized according to their exposure in the existing top-$K$ retrieval results, following the positional effect of Finding~2. This baseline uses retrieval position but does not inspect the new embedding.

    \item \textbf{YW-hist.}
    Updates with stronger prior substantive-change history are prioritized, following the temporal persistence of Finding~3.

    \item \textbf{True-$\delta$ Oracle.}
    Updates are ranked by their true embedding displacement after the new representation has been computed. This baseline violates the deployment information boundary and is therefore used only as an oracle reference.
\end{enumerate}

The first four policies are compatible with the benchmark's pre-embedding information boundary. True-$\delta$ is deliberately non-deployable: its role is to estimate how much improvement would remain available if future representation movement were known perfectly.

We additionally evaluate a character-feature LightGBM model as a \emph{diagnostic learned baseline}. It is not counted among the five reference policies above; its purpose is to test whether a lightweight trained predictor automatically dominates simple pre-embedding signals.

\subsection{Benchmark Characterization}

Table~\ref{tab:cipbench-leaderboard} reports per-repository Track~B reference results at the two representative budgets.

\begin{table*}[t]
\caption{CIPBench Track~B reference results (lower $\Hurt@B$ is better).}
\label{tab:cipbench-leaderboard}
\centering
\scriptsize
\setlength{\tabcolsep}{3.2pt}
\begin{tabular}{llcrrrrr}
\toprule
\textbf{Repo} & \textbf{Lang.} & \textbf{$B$} &
\textbf{Random} & \textbf{Recency} &
\textbf{$k$-occ} & \textbf{YW-hist} &
\textbf{True-$\delta$}\\
\midrule
flask & Py & 30 & .039 & .070 & .078 & .016 & .016\\
      &    & 50 & .023 & .062 & .047 & .016 & .008\\
aiohttp & Py & 30 & .054 & .065 & .011 & .022 & .022\\
        &    & 50 & .054 & .022 & .011 & .011 & .000\\
pylint & Py & 30 & .027 & .027 & .022 & .013 & .018\\
       &    & 50 & .022 & .013 & .022 & .009 & .013\\
requests & Py & 30 & .040 & .040 & .020 & .033 & .020\\
         &    & 50 & .026 & .033 & .000 & .033 & .013\\
click & Py & 30 & .036 & .024 & .012 & .024 & .012\\
      &    & 50 & .036 & .024 & .012 & .012 & .012\\
cobra & Go & 30 & .096 & .143 & .125 & .054 & .042\\
      &    & 50 & .063 & .110 & .075 & .036 & .024\\
echo & Go & 30 & .171 & .129 & .150 & .086 & .046\\
     &    & 50 & .129 & .103 & .103 & .065 & .025\\
gin & Go & 30 & .114 & .127 & .118 & .078 & .062\\
    &    & 50 & .107 & .116 & .098 & .053 & .027\\
lodash & JS & 30 & .098 & .098 & .027 & .030 & .030\\
       &    & 50 & .049 & .102 & .023 & .011 & .027\\
express & JS & 30 & .054 & .041 & .041 & .041 & .041\\
        &    & 50 & .041 & .027 & .041 & .068 & .041\\
\bottomrule
\end{tabular}
\end{table*}

The results exhibit substantial variation across repositories. No deployable single-signal baseline dominates in every setting. In particular, Recency underperforms Random on average at both budgets ($B{=}30$: .076 vs.\ .073; $B{=}50$: .061 vs.\ .055): the issue-driven stream concentrates recent updates at the tails of multi-update issues, so freshness alone selects the smallest, least representative steps. YW-hist performs particularly well in several repositories, whereas $k$-occ is stronger in others.

Table~\ref{tab:cipbench-mean} summarizes mean harm across the 10 repositories.

\begin{table}[htbp]
\caption{Mean Track~B harm across all 10 repositories. Differences quoted in the text are computed from unrounded per-repository values and may therefore differ from the quotient of the displayed figures (e.g., YW-hist improves on Recency by 3.67pp, which the rounded means give as 3.6pp).}
\label{tab:cipbench-mean}
\centering
\small
\begin{tabular}{crrrrr}
\toprule
$B$ & Random & Recency & $k$-occ & YW-hist & True-$\delta$\\
\midrule
30\% & .073 & .076 & .060 & \textbf{.040} & .031\\
50\% & .055 & .061 & .043 & \textbf{.031} & .019\\
\bottomrule
\end{tabular}
\end{table}

At $B=30\%$, YW-hist is the strongest deployable single-signal baseline on average, reducing mean harm by 3.3 percentage points relative to Random and by 3.7 points relative to Recency. A substantial gap also remains between Recency and the True-$\delta$ oracle: 4.5 percentage points at $B=30\%$ and 4.2 points at $B=50\%$.

Workload heterogeneity is visible at the language level as well
(Appendix~\ref{app:robustness}). Python is already comparatively close to the oracle
under these baselines, while the Go repositories leave substantially more room for
improvement, and in the two JavaScript repositories $k$-occ has lower mean harm than
YW-hist. These differences reinforce the observation from Section~\ref{sec:empirical}
that the usefulness of individual signals depends strongly on repository workload.

The diagnostic character-feature LightGBM baseline obtains mean harm of approximately .058 at $B=30\%$. It slightly improves on $k$-occ but remains worse than YW-hist. Thus, under the same pre-embedding information constraint, adding a trained character-level predictor does not by itself eliminate the advantage of simple historical signals.

Overall, the reference results establish two properties of CIPBench. First, the benchmark separates maintenance policies that exploit different kinds of pre-embedding information into clearly distinct harm levels. Second, a substantial gap remains between the strongest deployable baseline and an oracle with access to future embedding displacement. Whether that gap can be closed by combining the complementary signals of Section~\ref{sec:empirical} is the question Section~\ref{sec:trackb} takes up; Section~\ref{sec:tracka} then evaluates the filtering track.

\section{Track A: Conservative Filtering}\label{sec:tracka}

Track~A asks whether a changed unit can be skipped outright, before any ranking
question arises. Section~\ref{subsec:counterexamples} supplies two candidate rules: the
$A\rightarrow B\rightarrow A$ reversions, in which a unit's text returns to the state
the index already represents, and the docstring-only edits, the strongest observed
instance of ``embedding moves, retrieval does not.'' Both are cheap to detect before
re-embedding. This section measures what they save and what they cost.

\subsection{Rules and Measurement}

We evaluate two rules over the 2{,}366 units of all 10 repositories, using each unit's change history to decide:

\begin{enumerate}[leftmargin=*]
    \item \textbf{Hash-identity skip.} Skip a unit whose text at the refresh point is
    hash-identical to the text represented by the currently indexed vector
    (Section~\ref{subsec:funnel}).
    \item \textbf{Python docstring skip.} Additionally skip a unit whose accumulated
    changes are confined to Python docstrings
    (Section~\ref{subsec:counterexamples}).
\end{enumerate}

Both rules are stated over units because that is the decision object, but the evidence
they consult is update-level: the first compares the unit's current text against what
the index holds, and the second inspects the change types of the updates that produced
it. A unit qualifies for the second rule only if \emph{every} update in its accumulated
history is docstring-only; a single implementation change disqualifies it.

We evaluate both rules with the Track~A metrics of Section~\ref{sec:cipbench},
$\mathrm{Savings}$ and $\mathrm{MissRate}$, writing $S_{\mathrm{reembed}}$ for the set
there denoted $S$ to keep it distinct from the set a rule \emph{skips}.

We report the two rules separately rather than pooled. Hash-identical skips are
non-impacting by construction---identical text yields an identical embedding, which
cannot change any ranking---so they contribute zero misses by definition. Pooling them
with the docstring rule would divide the docstring rule's misses by a denominator
inflated with cases that cannot miss, overstating its safety.

\subsection{Results}

\begin{figure}[t]
\centering
\includegraphics[width=0.6\columnwidth]{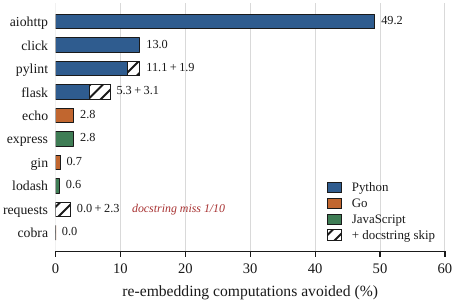}
\caption{Per-repository savings from conservative filtering. Hash-identical skipping is safe everywhere but heavily skewed, reaching 49.2\% in aiohttp where the reversion phenomenon is concentrated. The docstring rule adds savings in three Python repositories---pylint, flask and requests---and is the only rule to record a miss, in requests.}
\label{fig:tracka-savings}
\end{figure}

Table~\ref{tab:tracka} reports savings and miss rate per repository at each level;
Figure~\ref{fig:tracka-savings} shows the savings distribution.

\begin{table*}[t]
\caption{Track~A filter evaluation. ``Miss'' is the fraction of a repository's impacted units that the rule skips, following the definition in Section~\ref{sec:cipbench}. Hash-identity misses are zero by construction. The single docstring miss, in requests, is one unit out of that repository's ten impacted units---and out of the one unit to which the rule applies there.}
\label{tab:tracka}
\centering
\scriptsize
\setlength{\tabcolsep}{4pt}
\begin{tabular}{llrrrr}
\toprule
\textbf{Repo} & \textbf{Lang.} &
\textbf{hash-skip savings} & \textbf{hash-skip miss} &
\textbf{+docstring savings} & \textbf{+docstring miss}\\
\midrule
aiohttp & Py & 49.2\% & 0 & 49.2\% & 0\\
click & Py & 13.0\% & 0 & 13.0\% & 0\\
pylint & Py & 11.1\% & 0 & 13.0\% & 0\\
flask & Py & 5.3\% & 0 & 8.4\% & 0\\
echo & Go & 2.8\% & 0 & 2.8\% & 0\\
express & JS & 2.8\% & 0 & 2.8\% & 0\\
gin & Go & 0.7\% & 0 & 0.7\% & 0\\
lodash & JS & 0.6\% & 0 & 0.6\% & 0\\
requests & Py & 0.0\% & 0 & 2.3\% & \textbf{0.10} (1/10)\\
cobra & Go & 0.0\% & 0 & 0.0\% & 0\\
\bottomrule
\end{tabular}
\end{table*}

The hash-identity rule has a zero miss rate in every repository, as its construction
requires, and its savings are heavily skewed: 49.2\% in aiohttp, where the reversion
phenomenon is concentrated, but under 3\% in six of the ten repositories and exactly
zero in cobra and requests. The rule is worth deploying because it never costs
anything, not because it reliably saves much.

The docstring rule is safe in four of five Python repositories but \textbf{misses one
impacted unit in requests}, the only miss across all 10 repositories. The reported rate
of 0.10 follows the definition above: one miss over requests' ten impacted units. Read
against the rule's own candidates it is far worse, and worth stating plainly---the rule
applies to exactly one unit in requests, and that unit is impacted. Its 2.3\% saving and
its miss are the same unit.

That is the reason we do not present the rule as universal. It is also why the
repository-level miss rate understates the risk: a rule with one candidate and one miss
looks like a 10\% failure against the repository and a total failure against itself, and
only the second number describes what the rule does.

Extending the filter further is unsafe on the current evidence. Comment-only changes
miss 2/19 impacted units in flask and 2/8 in click, confirming the comment-sensitivity
counterexample of Section~\ref{subsec:counterexamples}. Go string-literal changes miss
1/87 in cobra and 1/165 in gin. Neither class should be hard-filtered.

\subsection{A Label-Free Deployment Guard}

The requests miss creates a design problem. Hard-coding a repository exclusion would
overfit the benchmark, and a deployed system cannot consult retrieval-impact labels---
computing them requires exactly the embeddings the filter exists to avoid. We
therefore express the rule as a guard that a deployment can evaluate from its own
observations:

\begin{quote}
Enable the docstring filter for a repository only after periodic probing has
accumulated at least $N=20$ observed docstring-only updates with zero measured impact;
disable it permanently at the first observed miss.
\end{quote}

Probing re-embeds a small random sample of filtered units and measures their true
impact, so the guard is self-supervising: it pays a bounded sampling cost to maintain
the evidence its own safety depends on. $N=20$ is a deployment parameter and we state it
explicitly, because leaving it unspecified would reintroduce the overfitting objection in
a vaguer form.

We must be clear about what the benchmark can say regarding this threshold, which is
less than we would like. The docstring rule applies to eight units in total across the
ten repositories---four in flask, three in pylint, one in requests---so no repository
here accumulates anything close to twenty clean observations within the observed window.
On this benchmark the guard would therefore enable the rule nowhere; it excludes requests
not by tripping on the miss but by never reaching the evidence threshold at all. The
guard is a deployment mechanism whose counts accumulate over a production stream far
longer than an issue-linked evolution window, and the benchmark can demonstrate its
logic but cannot exercise it.

This bounds the Track~A result. The hash-identity rule carries essentially all of the
measured savings and is safe by construction; the docstring rule rests on eight
candidate units, one of which is impacted, and should be read as a design sketch
supported by the counterexample analysis of Section~\ref{subsec:counterexamples} rather
than as a validated rule.

\subsection{Deployment Cost and Interface}\label{subsec:deployment}

The filter is the first stage of a per-unit decision path whose second stage, scoring,
is the subject of Section~\ref{sec:trackb}. We give the whole path here
because the two stages share one traversal of the update.
Algorithm~\ref{alg:should-reembed} states it. The decision respects the CIPBench
information boundary throughout: it consults the indexed text, the updated source
text, prior change history, and the old index, and never the updated embedding, true
displacement, or retrieval-impact label.

\begin{algorithm}[t]
\caption{Priority computation for one changed unit.}
\label{alg:should-reembed}
\begin{algorithmic}[1]

\Require Changed unit $u$; new text $x_{\mathrm{new}}$; indexed state;
         weights $w_0,w_1,w_2,w_3$
\Ensure \textsc{Skip} or a priority score

\If{$\Call{Hash}{x_{\mathrm{new}}}
    = \Call{IndexedTextHash}{u}$}
    \State \Return \textsc{Skip}
\EndIf

\State $x_{\mathrm{old}} \gets \Call{IndexedText}{u}$
    \Comment{text the index currently represents}
\State $ct \gets
    \Call{ClassifyTreeSitter}{x_{\mathrm{old}},x_{\mathrm{new}}}$

\If{$ct = \mathrm{PythonDocstring}$
    \textbf{ and }
    $\Call{DocstringFilterEnabled}{repo}$}
    \State \Return \textsc{Skip}
\EndIf

\State $k_{\mathrm{occ}}
    \gets \Call{ComputeKOcc}{u,K=10}$

\State $s_{\mathrm{nn}}
    \gets \Call{NNSim}{u}$

\State $n_{\mathrm{sub}} \gets \Call{NSubCount}{u}$
    \Comment{substantive updates in $u$'s history}
\State $score \gets
    w_0 \cdot \mathbf{1}[n_{\mathrm{sub}} \ge 1]
    + w_1 \cdot \log(1+k_{\mathrm{occ}})$
\Statex \hspace{\algorithmicindent}
    $\phantom{score \gets{}}
    + w_2 \cdot \log(1+n_{\mathrm{sub}})
    + w_3 \cdot s_{\mathrm{nn}}$

\State \Return $score$

\end{algorithmic}
\end{algorithm}

The two filter stages are inexpensive by construction. A hash comparison is $O(1)$ in
the length of the unit, and tree-sitter classification is millisecond-scale on CPU, so
an update that the filter skips leaves the path having consumed neither a GPU nor an
index traversal---which is the entire point of placing the filter first.

The scoring stage is where the cost sits. Historical counts are $O(1)$ once maintained
in memory, but computing $k$-occ requires a query--unit ranking pass over the old index
at $O(n_q n_u \log n_u)$, by a wide margin the most expensive step. Whether a deployment
must pay it depends on the weight its repository's cross-validation assigns to $k$-occ:
Section~\ref{sec:trackb} reports that weight as zero in eight of the ten repositories,
where the pass can be omitted without changing any decision, and small but non-zero in
the remaining two.

\textbf{Result 1.}
\emph{Conservative filtering avoids measurable computation at measurable risk. Skipping
hash-identical units is safe in every repository by construction and saves up to 49.2\%,
which is the whole of the measured benefit. The Python docstring rule applies to eight
units across the benchmark and misses one of them, too few to validate; we give a
label-free probing guard for deploying it safely but cannot exercise that guard here.
Comment-only and string-literal extensions are unsafe on the current evidence.}
\section{Track B: History Alone Captures the Available Predictability}\label{sec:trackb}

Filtering removes updates that provably cannot matter. Track~B asks the harder
question that remains: among the units that survive the filter, which should be
re-embedded first when only a fraction of them can be?

Section~\ref{sec:empirical} leaves a prior worth testing here. Four signals---syntactic
substantiality, retrieval exposure ($k$-occ), substantive-change history ($n_{sub}$),
and old-index neighborhood similarity (\texttt{nn\_sim})---are the inexpensive signals
available at the boundary. Three of them discriminate impacted units and
\texttt{nn\_sim} does not (Finding~5); we retain it so that the combination is tested
with everything the boundary permits rather than with a set pre-filtered on the outcome.
Five of the six pairings are weakly associated
(Appendix~\ref{app:signal-independence}); the exception, change type versus history, is
the pair Section~\ref{subsec:policy-design} shows to be related by construction. Standard practice would take that as
sufficient reason to combine them. This section asks whether the combination actually
improves the re-embedding decision, measured against the strongest \emph{deployable
single signal} rather than against an arbitrary baseline.

Prediction works here, and one signal does all of it. Section~\ref{sec:cipbench}
already establishes that the history baseline YW-hist reduces mean harm by 3.3
percentage points relative to Random and 3.7 relative to Recency---a substantial and
deployable margin under the pre-embedding boundary. What the combination adds to that
margin is nothing. On the complete 20-point operating grid a weighted combination of
all four signals is statistically indistinguishable from YW-hist alone, and the
conclusion survives every perturbation of the score we could apply. The available
predictability is real; it is simply exhausted by a single cheap signal.

\subsection{Policy, Baselines, and Reporting}\label{subsec:policy-design}

\subsubsection{The weighted policy}

For every update that survives the conservative filter of
Section~\ref{sec:tracka}, the weighted score for unit $u$ is
\begin{align}
\label{eq:score}
\mathrm{score}(u)={}&
w_0\,\mathbf{1}[\mathrm{substantive}(u)]\notag\\
&+w_1\log(1+k\text{-occ}(u))\notag\\
&+w_2\log(1+n_{sub}(u))
+w_3\,\mathrm{nn\_sim}(u),
\end{align}
with the weighted terms normalized so that $w_1+w_2+w_3=1$.

One relationship between the terms should be stated at the outset, because it governs
what follows. At the unit level a unit is flagged substantive exactly when it has at
least one substantive update in its history, so $\mathbf{1}[\mathrm{substantive}(u)]$
and $n_{sub}(u)$ are not independent quantities: the flag is $n_{sub}\ge1$. We verified
this holds without exception across all 2{,}366 units. Equation~\ref{eq:score} therefore
combines a count with a thresholded copy of that same count, and the question this
section asks is in part whether the threshold adds anything to the count. We select the
weights by
leave-one-repository-out cross-validation over a fixed 64-point grid; when repository
$r$ is evaluated, no outcome from $r$ contributes to its weights, so the procedure
yields one weight vector per repository.

The selected vectors are strikingly uniform. Eight of the ten repositories select
$(w_1,w_2,w_3)=(0,\,0.667,\,0.333)$, assigning \emph{zero} weight to retrieval
exposure; the exceptions are flask, which selects $(0.167,\,0.667,\,0.167)$, and
pylint, which selects $(0.2,\,0.8,\,0)$. Cross-validation therefore discards $k$-occ
in most repositories and retains a small amount of it in two, which is consistent with
the weak and repository-dependent discrimination reported in Finding~5. We treat $w_0$
as a tunable coefficient and sweep it in Section~\ref{subsec:w0} so that the result
cannot be attributed to the choice of its value.

Where the selection is $w_1=0$ it lies on the boundary of the coarse grid, which has no
sample between zero and its smallest positive value. A fine grid near $w_1{=}0$ (step
0.005) confirms that harm is flat from $w_1{=}0$ through approximately 0.03 in cobra and
0.05 in echo before worsening, so those selections sit at the boundary of a flat optimum
rather than at a unique minimum. The grid is therefore too coarse to distinguish a zero
weight on $k$-occ from a small positive one, which is also why flask and pylint can
select $w_1=0.167$ and $0.2$ without contradicting the eight repositories that select
zero.

\subsubsection{Reference baselines}

Beyond the five baselines of Section~\ref{sec:cipbench} (Random, Recency, $k$-occ,
YW-hist, True-$\delta$), we add two rankers that isolate individual contributions:

\begin{enumerate}[leftmargin=*]
    \item \textbf{Substantive-only}, ranking by the substantive flag alone with stream
    order as tie-breaker, which tests whether Eq.~\ref{eq:score} reduces to the
    substantive partition at the selection level.
    \item \textbf{YW-hist $+$ $k$-occ}, the two-signal combination
    $w_1\log(1+k\text{-occ})+w_2\log(1+n_{sub})$ with weights selected by the same
    procedure, which tests whether adding the orthogonal exposure signal to the
    strongest history signal helps.
\end{enumerate}

\subsubsection{Reporting rule}

CIPBench defines budgets $B\in\{10,\ldots,100\}\%$. We evaluate the full
repository--budget grid at $B\in\{30,50\}$---all 20 operating points, none omitted.
Table~\ref{tab:strategy-full} reports the complete grid.

\begin{table*}[t]
\caption{Full Track~B operating-point grid (lower harm is better). $\Delta$ is Weighted $-$ YW-hist harm; positive means the weighted policy is worse.}
\label{tab:strategy-full}
\centering
\scriptsize
\setlength{\tabcolsep}{4pt}
\begin{tabular}{llcrrrrr}
\toprule
\textbf{Repo} & \textbf{Lang.} & \textbf{$B$} &
\textbf{Recency} & \textbf{YW-hist} & \textbf{Sub-only} &
\textbf{Weighted} & \textbf{$\Delta$}\\
\midrule
flask & Py & 30 & .070 & .016 & .055 & .039 & +2.34pp\\
flask & Py & 50 & .062 & .016 & .016 & .039 & +2.34pp\\
aiohttp & Py & 30 & .065 & .022 & .054 & .022 & 0\\
aiohttp & Py & 50 & .022 & .011 & .054 & .011 & 0\\
pylint & Py & 30 & .027 & .013 & .045 & .018 & +0.45pp\\
pylint & Py & 50 & .013 & .009 & .031 & .018 & +0.90pp\\
requests & Py & 30 & .040 & .033 & .033 & .033 & 0\\
requests & Py & 50 & .033 & .033 & .033 & .033 & 0\\
click & Py & 30 & .024 & .024 & .024 & .024 & 0\\
click & Py & 50 & .024 & .012 & .012 & .012 & 0\\
cobra & Go & 30 & .143 & .054 & .096 & .051 & $-0.30$pp\\
cobra & Go & 50 & .110 & .036 & .051 & .036 & 0\\
echo & Go & 30 & .129 & .086 & .143 & .086 & 0\\
echo & Go & 50 & .103 & .065 & .133 & .065 & 0\\
gin & Go & 30 & .127 & .078 & .118 & .078 & 0\\
gin & Go & 50 & .116 & .053 & .083 & .054 & +0.18pp\\
lodash & JS & 30 & .098 & .030 & .049 & .027 & $-0.38$pp\\
lodash & JS & 50 & .102 & .011 & .019 & .015 & +0.38pp\\
express & JS & 30 & .041 & .041 & .054 & .041 & 0\\
express & JS & 50 & .027 & .068 & .054 & .068 & 0\\
\midrule
\multicolumn{8}{l}{\footnotesize Mean $\Delta=+0.30$pp, median $\Delta=0$, 12 ties, 2 wins, 6 losses.}\\
\bottomrule
\end{tabular}
\end{table*}

\subsection{The Combination Is Equivalent to History Alone}\label{subsec:trackb-results}

\begin{figure}[t]
\centering
\includegraphics[width=0.5\columnwidth]{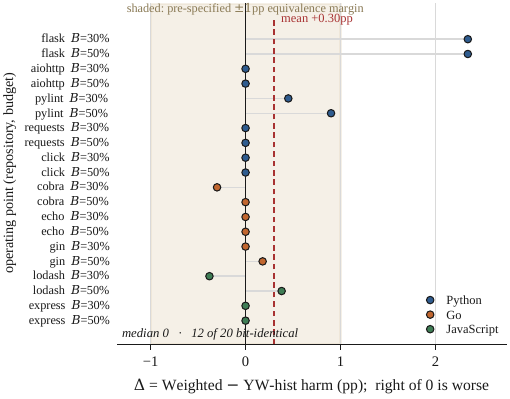}
\caption{Per-point difference between the weighted policy and YW-hist at each of the 20 operating points. Twelve points are exactly zero. The mean is driven almost entirely by flask, whose two points are the only ones outside the pre-specified $\pm1$pp margin.}
\label{fig:delta-equivalence}
\end{figure}

Table~\ref{tab:equiv-stats} collects the comparison and
Figure~\ref{fig:delta-equivalence} shows its per-point distribution. Two features of
that distribution carry the claim. Twelve of the 20 points are bit-identical, so on most
of the grid the two policies are not merely close but indistinguishable on the metric.
Identical harm does not by itself establish identical selection---$\mathrm{Hurt@B}$
counts degraded queries, and two different selections can produce the same count---so
Section~\ref{subsec:w0} compares selected sets directly where that stronger claim is
needed. And the positive mean is almost entirely one repository: flask contributes $+2.34$pp
at both budgets, four fifths of the total, leaving the other 18 points averaging
$+0.07$pp. The median difference is zero.

\begin{table}[htbp]
\caption{Equivalence statistics between the weighted policy and YW-hist over the 20-point grid.}
\label{tab:equiv-stats}
\centering
\small
\begin{tabular}{lr}
\toprule
\textbf{Statistic} & \textbf{Value}\\
\midrule
Mean $\Delta$ (Weighted $-$ YW-hist) & $+0.30$pp\\
Median $\Delta$ & 0\\
Bit-identical ties / 20 & 12\\
Wins / Losses & 2 / 6\\
Repository-clustered bootstrap 95\% CI & $[-0.02, +0.79]$pp\\
TOST ($\pm1$pp margin) & $p=2.4\times10^{-4}$\\
Simulated coverage ($\sigma{=}0.01/0.02$) & 0.91 / 0.86\\
CI width ratio ($2\times\sigma$) & 1.90\\
\bottomrule
\end{tabular}
\end{table}

Two qualifications bound this claim. The $\pm1$pp margin is absolute, so on
repositories with very low harm---aiohttp ranges from .011 to .022---it is roughly
half the quantity being compared; equivalence would be looser under a relative margin.
And the clustered bootstrap undercovers at 10 clusters: a synthetic validation with
known between-repository variance measures coverage at 0.91 ($\sigma{=}0.01$) and 0.86
($\sigma{=}0.02$) against a nominal 0.95, with CI width scaling near-linearly (ratio
1.90 for a doubled $\sigma$).

The weighted policy does improve on Recency by 3.04pp, but Recency is itself beaten by
Random (Section~\ref{sec:cipbench}), so that comparison measures nothing. We treat
YW-hist as the reference throughout.

\textbf{Result 2.}
\emph{Combining four pre-embedding signals is statistically equivalent to using the
single strongest one, within $\pm1$pp over the full 20-point grid
(Table~\ref{tab:equiv-stats}). Twelve of 20 operating points produce bit-identical
harm.}

\subsection{The Result Does Not Depend on the Substantive Weight}\label{subsec:w0}

A fixed large coefficient on the substantive term would guarantee that substantive
updates outrank non-substantive ones, inviting the objection that Result~2 is an
artifact of parameterization. We therefore sweep $w_0\in\{0,\,0.94,\,3.38,\,5.06,\,10\}$, re-selecting $w_1,w_2,w_3$
at each value. Every summary statistic is unchanged at every value: mean $\Delta=+0.30$pp,
median $0$, 12 ties, and the same per-repository weight selections
(Appendix~\ref{app:robustness}).

Exact invariance across structurally different score functions calls for a check on what
is actually happening, and the answer is set identity: the selected sets $S_B$ at
$w_0{=}0$ and $w_0{=}10$ coincide at all 20 operating points (Jaccard $1.0$), so the
harm must be bit-identical. At these budgets the substantive tier of ranked units
exceeds the budget in every repository (Section~\ref{subsec:partition}), which is
consistent with the flag having no room to act. We report the invariance as observed
rather than claiming the tier condition as its cause; separating the two would require
budgets at which the tier no longer covers the selection, and those lie well above the
operating range studied here.

A positive control confirms that the score does respond to $w_0$: inverting the sign of
the substantive term at pylint $B{=}30$ changes the selected set by as much as the
repository's composition permits.\footnote{At $w_0{=}{-}100$ the selection is pylint's 12
non-substantive units plus the top-36 substantive units; at $w_0{=}{+}100$ it is the
top-48 substantive units. Twelve of 48 differ---the maximum available, since only 12 of
pylint's 162 ranked units are non-substantive, so the budget cannot be filled from that
tier alone.}

The identity noted in Section~\ref{subsec:policy-design} explains why, and does so
exactly rather than empirically. Because a substantive unit has $n_{sub}\ge1$ and a
non-substantive unit has $n_{sub}=0$, the history term is bounded below by
$w_2\log 2$ on the substantive side and contributes nothing on the other. In the eight
repositories that select $(w_1,w_2,w_3)=(0,\,0.667,\,0.333)$ this gives
$T(u_s)\ge0.462$ while $T(u_{ns})\le w_3\max(\mathrm{nn\_sim})\le0.333$, so $T$ orders
every substantive unit above every non-substantive one and
$w_0\mathbf{1}[\mathrm{substantive}]$ cannot reorder anything for any $w_0\ge0$. The
invariance in those repositories is an algebraic consequence of the score's
construction, not a property of the data.

The two repositories that retain a non-zero weight on retrieval exposure are the
exception, and the same algebra says where. For pylint, $T(u_s)\ge0.555$ while a
non-substantive unit reaches that level once $k\text{-occ}>15$; for flask the threshold
is $k\text{-occ}>5$. Separation therefore holds in those repositories only while no
sufficiently exposed non-substantive unit exists near the budget cutoff, which is a
contingent fact rather than a guarantee. Section~\ref{subsec:partition} finds exactly
this: the flag is inert everywhere at budgets up to 70\%, and becomes active only at
higher budgets and only in flask and pylint.

\textbf{Result 3.}
\emph{Result~2 is not a parameterization artifact. The selected sets at $w_0{=}0$ and
$w_0{=}10$ coincide at all 20 operating points, so the reported harm is invariant to the
substantive weight over $[0,10]$, and a positive control confirms the score does respond
to $w_0$. Section~\ref{subsec:partition} shows the invariance extends to every budget at or
below 70\%.}

\subsection{Why the Flag Is Inert, and Over What Range}\label{subsec:partition}

Substantive-only and Weighted differ in harm at 16 of 20 points, sometimes by a factor
of two or more (echo $B{=}50$: .133 vs.\ .065; aiohttp $B{=}50$: .054 vs.\ .011). That
looks at first like the substantive partition failing, but comparing selected sets
rather than harm shows the opposite.

Every repository's substantive fraction of \emph{ranked units}---a different quantity
from the update-level temperature of Section~\ref{subsec:context}---exceeds 87\%, from
87.0\% in click to 99.1\% in gin. At both $B{=}30\%$ and $B{=}50\%$ the substantive tier
is therefore larger than the budget in every repository, so every selected unit is
substantive whatever the flag's value and the flag cannot discriminate at all. What
differs between the two policies is the within-tier ordering---stream index versus the
weighted terms---and that ordering changes the selected set at 19 of 20 points, with
Jaccard overlap between 0.12 and 0.45. The large harm differences in
Table~\ref{tab:strategy-full} follow from those set differences, not from any failure of
the partition.

Reporting this at two budgets invites the objection that we evaluated where the signals
happened not to matter. We therefore repeat the comparison over
$B\in\{5,10,20,\ldots,95,100\}\%$, giving 130 repository--budget points with each
repository's own cross-validated weights.

The picture does not change with the budget. Across all 130 points the mean difference
is $+0.059$pp with a median of $0$, and 71 points are bit-identical.
Table~\ref{tab:budget-sweep} gives the per-budget means, whose sign alternates without
pattern across the thirteen levels---what one expects from noise around zero rather than
from a budget-dependent effect.

We describe this as consistency with Result~2 rather than as equivalence, because it is
weaker evidence of a different kind. The pre-specified margin, the clustered bootstrap
and the TOST apply to the 20-point grid, where the operating points were fixed in
advance; the sweep is a check that no budget reverses the conclusion, and 17 of its 130
points do exceed one percentage point individually. What it rules out is a
budget-dependent effect that the two reported budgets happened to miss, which is the
objection it was run to answer.

\begin{table}[htbp]
\caption{Mean $\Delta$ (Weighted $-$ YW-hist, pp) over the 10 repositories at each budget, with the number of bit-identical points and the number at which the substantive flag changes the selected set.}
\label{tab:budget-sweep}
\centering
\small
\begin{tabular}{crrr}
\toprule
$B$ & \textbf{Mean $\Delta$ (pp)} & \textbf{Ties / 10} & \textbf{Flag active / 10}\\
\midrule
5\%   & $-0.086$ & 7 & 0\\
10\%  & $-0.213$ & 5 & 0\\
20\%  & $+0.573$ & 2 & 0\\
30\%  & $+0.211$ & 6 & 0\\
40\%  & $+0.268$ & 4 & 0\\
50\%  & $+0.380$ & 6 & 0\\
60\%  & $+0.104$ & 6 & 0\\
70\%  & $-0.005$ & 4 & 0\\
80\%  & $-0.184$ & 5 & 1\\
85\%  & $-0.290$ & 5 & 2\\
90\%  & $-0.124$ & 5 & 2\\
95\%  & $+0.128$ & 6 & 1\\
100\% & $+0.000$ & 10 & 0\\
\bottomrule
\end{tabular}
\end{table}

The sweep also bounds the invariance of Result~3. Comparing the sets selected under
$w_0{=}0$ and $w_0{=}10$ at each point, they are identical at every budget at or below
70\%. The flag does change the selection at six points, all at $B\ge80\%$ and all in
flask and pylint---the two repositories whose cross-validated weights give $k$-occ a
non-zero coefficient. Where $w_1=0$ the ranking term is a function of history and
neighbourhood similarity alone, both aligned with substantiality, and the flag has
nothing left to reorder; where $w_1>0$ the term carries retrieval exposure, which is not
so aligned, and the flag can act. Even there it acts only when four fifths or more of
all updates are re-embedded, and the largest such perturbation moves the selected set by
under 4\% (Jaccard $\ge0.968$).

\textbf{Result 4.}
\emph{The substantive flag is inert at every budget a deployment would impose. The
substantive tier of ranked units (87.0\%--99.1\%) exceeds the budget at all reported
operating points, and over the full range the flag changes the selected set at no point
with $B\le70\%$ and at six points above it, confined to the two repositories in which
cross-validation retains a non-zero weight on retrieval exposure. Within-tier ordering,
not the flag, determines the selection: $S_B$(Substantive-only) differs from
$S_B$(Weighted) at 19 of 20 points. Extending the comparison to 130 repository--budget
points leaves the picture unchanged (mean $+0.059$pp, median $0$, 71 bit-identical), so
Result~2 is not an artifact of the two budgets reported.}

\subsection{Adding the Orthogonal Signal Does Not Help Either}\label{subsec:yw-kocc}

Result~2 could be read narrowly, as a fact about one particular four-term score. The
sharper test is the two-signal combination that the Section~\ref{sec:empirical}
evidence most directly recommends. $k$-occ is only weakly associated with $n_{sub}$
(language-level Kendall $\tau=0.13$, $0.11$, and $0.24$ for Python, Go, and
JavaScript), so if orthogonality translated into decision-level gain anywhere, it
should be here.

It does not. Over the 20 points the two-signal ranker ties YW-hist on 14, is worse on 5
(flask $B{=}30/50$, pylint $B{=}30/50$, gin $B{=}30$), and better on 1 (gin $B{=}50$,
$-0.36$pp), for a mean difference of $+0.21$pp. Adding retrieval exposure to a history
ranker does not reduce harm.

\textbf{Result 5.}
\emph{Signal-level independence does not imply decision-level gain. $k$-occ is
near-orthogonal to $n_{sub}$ and individually discriminative, yet adding it to a
history ranker changes no aggregate outcome (14/20 ties; mean $+0.21$pp worse).}

\subsection{Why the Displacement Oracle Is Not a Ceiling}\label{subsec:oracle}

Two operating points look striking in isolation: at aiohttp $B{=}30$ the weighted
policy matches the True-$\delta$ oracle (.022 vs.\ .022), and at lodash $B{=}50$ it
beats it (.015 vs.\ .027). Neither is a success of the combination---YW-hist attains
exactly the same harm at both points.

The observation still matters, because it confirms Finding~2 at the policy level.
Displacement alone does not characterize retrieval impact, so an oracle that ranks by
displacement is not a retrieval-aware scheduler and its harm is not a lower bound on
what a deployable policy can achieve. This is a property of the task, and it is one
reason the Recency-to-oracle gap reported in Section~\ref{sec:cipbench} should not be
read as headroom that a better ranker is guaranteed to capture.

\subsection{Summary}\label{subsec:trackb-summary}

A single cheap historical signal captures essentially all of the predictability
available before re-embedding. Three further signals---two of them discriminative, all largely unassociated with one
another, and combined under cross-validated weights---add nothing to it. The result holds across the full budget range, from 5\% to 100\%, and is invariant to
the substantive weight. At the budgets that matter the substantive tier exceeds the
budget in every repository, leaving the syntactic flag no room to discriminate, and
history carries the decision.

Read together with Section~\ref{sec:tracka}, the two tracks fail and succeed for
opposite reasons. Ordering saturates because the discriminative work is already done by
a single signal. Filtering succeeds because it exploits a structural property---textual
identity between the new state and the indexed one---that no ranking signal captures,
and that certifies non-impact rather than predicting it. Section~\ref{sec:discussion}
takes up what this contrast implies for how maintenance signals should be validated.

\section{Threats to Validity}\label{sec:threats}

Several aspects of the study affect how broadly its findings can be interpreted, including the choice of embedding model, change labeling, dataset construction, and repository coverage. We discuss six threats below, focusing on the scope of the conclusions they affect and the evidence available to assess them.

\paragraph{T1---Embedding-model dependence.}
The main analysis uses MiniLM-L6-v2, so some observed relationships may depend on the embedding model. We assess this by repeating the displacement and retrieval-impact measurements with MPNet-base-v2 and BGE-base-en-v1.5 across all 10 repositories.

Displacement rankings are highly consistent across the three models: over 30 repository--model comparisons the mean Spearman correlation is $\rho=0.905$ (range 0.82--0.97), with language-level means of 0.903, 0.901, and 0.915 for Python, Go, and JavaScript. Retrieval impact is not. The impacted sets produced by MPNet and BGE overlap only partially with those from MiniLM: mean Jaccard overlap 0.232 across 20 comparisons, ranging from 0.07 to 0.38. The relative displacement patterns therefore transfer well across the tested encoders, whereas the exact set of retrieval-impacting updates does not.

Two caveats bound the second measurement. The cross-model retrieval analysis differs slightly from the main pipeline for five repositories---flask, cobra, echo, gin, and lodash---where a stricter impact implementation produces impacted fractions 2--16 percentage points lower; the remaining five use matching definitions. This does not affect the main MiniLM results, but it means we can characterize the relative stability of displacement and the model sensitivity of impact without claiming exact transfer rates. Together these imply that the standard CIPBench impact labels are MiniLM-specific: the encoder is a parameter of the benchmark, and labels for a target encoder must be regenerated with the released pipeline rather than transferred.

\paragraph{T2---Change-label reliability.}
The empirical analysis relies on automatically assigned change categories. Section~\ref{subsec:empirical-setup} reports the validation results---Cohen's $\kappa=0.786$ between annotators~\cite{landis1977measurement} and 88.7\% classifier agreement with adjudicated labels; what matters for validity is the structure of the disagreement rather than its magnitude.

Disagreements concentrate on a small number of recurring classification boundaries rather than spreading uniformly. More than half (23/45) involve \texttt{lex\_substantive}. Agreement is also lower in cobra and lodash ($\kappa=.555$ and $.494$), where two specific ambiguities recur: Go template strings containing embedded DSL syntax, and local parsing failures in the large \texttt{runInContext} wrapper.

Adjudication assigns 31 disagreements to implementation changes, eight to signature changes, and six to string-literal-only changes; none remains in the \texttt{lex\_substantive} category. We subsequently clarified the annotation protocol so that local parsing failures with identifiable semantics are assigned to their semantic change class, while \texttt{lex\_substantive} is reserved for cases that cannot be classified reliably from the parsed structure. Following standard inter-rater-reliability practice~\cite{mchugh2012interrater}, we retain the original independently obtained $\kappa$ rather than recomputing agreement after this clarification.

Importantly, the main analyses use a broader substantive tier that combines \texttt{lex\_substantive} with implementation changes, so the principal findings are less sensitive to the fine-grained boundary responsible for most annotation disagreement.

\paragraph{T3---Issue-linked evolution streams.}
CIPBench follows issue-linked repository evolution rather than an unconstrained production commit stream. This choice provides a natural semantic anchor for retrieval queries, but it can also change the distribution of updates represented in the benchmark. Maintenance activities only weakly associated with issues---some dependency updates, routine refactorings, or automated changes---may be less visible in the resulting streams.

The current benchmark representation does not retain reliable author metadata, so bot-generated activity cannot be quantified separately. We do not rely on an assumed bot-contamination rate in any of the main analyses. The relevant generalization question is instead whether the observed temporal persistence, repository-level heterogeneity, and policy behavior remain similar under natural production commit streams; replication on bot-filtered commit histories would provide complementary evidence.

\paragraph{T4---Repository and language coverage.}
The benchmark contains five Python, three Go, and two JavaScript repositories, which vary substantially in size, change history, and retrieval behavior but remain 10 projects in three languages. Our conclusions should be read as cross-repository evidence within the studied ecosystems rather than as universal language-level laws. The qualification is sharpest for findings involving signatures, type annotations, comments, and string literals, whose roles differ across languages and coding conventions.

\paragraph{T5---Statistical multiplicity and repository-level sample size.}
Several parts of the analysis involve repeated statistical comparisons. For the three pairwise change-type comparisons underlying Table~\ref{tab:delta-by-change}, we use a Bonferroni-adjusted threshold of $\alpha'=0.017$; all three Mann--Whitney tests have $p<10^{-9}$ and remain significant after correction. The temporal-persistence analysis is reported separately for each repository, with 6/10 Fisher tests significant at $\alpha=0.05$, so we interpret it primarily through the magnitude and cross-repository pattern of the probability differences rather than through a pooled count of significant tests.

Repository-level meta-feature correlations (Appendix~\ref{app:meta-features}) serve a different purpose. They are computed over the 10 repositories to characterize possible sources of heterogeneity, not to establish population-level relationships, and no policy in this paper consults them.

The temporal-persistence analyses of Section~\ref{subsec:context} take the update as the unit of observation---both the first-update conditional probabilities and the consecutive-class persistence rates of Appendix~\ref{app:persistence}---even though a unit contributes several updates and may revert (e.g., aiohttp's $A\rightarrow B\rightarrow A$ annotation restorations, retained as real events). Their figures should be read as descriptive of the pooled update population rather than as exact inference over independent samples.

The remaining analyses are not affected in that way, because they are unit-level: displacement, retrieval impact, and three of the four signals are defined per unit, and change type is aggregated to the unit before use. Their dependence is of a different kind---a unit spanning several change classes is assigned only its most severe---which affects the class labels rather than the observation count.

% The Section~\ref{sec:trackb} equivalence claim rests on a repository-clustered bootstrap (1{,}000 resamples over the 10 repositories) and a two one-sided test with a pre-specified $\pm1$pp margin. 
The Section~\ref{sec:trackb} equivalence claim rests on a repository-clustered bootstrap(1{,}000 resamples over the 10 repositories) and the two one-sided tests
(TOST) procedure, using a pre-specified equivalence margin of $\pm 1$ pp.
With only 10 clusters, percentile bootstrap intervals are known to undercover: a synthetic-data validation with known between-repository variance measures coverage at 0.91 ($\sigma{=}0.01$) and 0.86 ($\sigma{=}0.02$), with CI width scaling near-linearly (ratio 1.90 for a $2\times$ standard deviation). We report the measured coverage rather than assuming the nominal 0.95. Two of the four signals entering Eq.~\ref{eq:score} are not independent by construction: the unit-level substantive flag is exactly $n_{sub}\ge1$, verified without exception across all 2{,}366 units. The equivalence reported in Section~\ref{sec:trackb} is therefore in part a statement about a score that combines a count with a thresholded copy of itself, and the reader should not read it as evidence that four unrelated signals fail to combine. The three genuinely distinct inputs are history, retrieval exposure, and neighbourhood similarity; Appendix~\ref{app:signal-independence} shows those to be weakly associated, and Section~\ref{subsec:yw-kocc} tests the strongest pair of them directly.

Separately, the weight grid limits what can be concluded about $k$-occ. Leave-one-repository-out selection returns $w_1=0$ in eight repositories and $0.167$ and $0.2$ in the remaining two, but the coarse 64-point grid has no sample between zero and its smallest positive value, and a fine sweep (step 0.005) finds harm flat from $w_1{=}0$ through roughly 0.03--0.05. The eight zero selections should therefore be read as ``no more than a small weight'' rather than as an exact zero, and the contrast between the two groups of repositories as a difference in degree rather than in kind.

Across the empirical analysis, we report effect sizes and cross-repository patterns alongside significance tests so that no conclusion depends on individual $p$ values alone.

\paragraph{T6---Filter boundary conditions and Track~A measurement.}
The Track~A evaluation of Section~\ref{sec:tracka} measures the two filter rules over all 2{,}366 units, so their savings and miss rates are observed rather than assumed. Three limits bound what that measurement establishes.

First, the docstring rule's evidence base is small in absolute terms. It changes the outcome in only three repositories---adding four skipped units in flask, three in pylint and one in requests---and the single miss we observe is that one requests unit. Expressed as the paper's MissRate it is one of requests' ten impacted units, or 0.10; expressed against the rule's own candidates in that repository it is one of one. A rule validated on tens of cases per repository should be deployed behind the probing guard we describe rather than enabled by default, and the guard's threshold ($N=20$) is an engineering choice calibrated against a single observed failure, not a value derived from a risk model.

Second, the hash-identity rule's zero miss rate is a construction guarantee, not an empirical one, and it should not be read as evidence about the docstring rule. We report the two rules separately for this reason. The reader should also note that MissRate is defined against a repository's impacted set, so a rule that fires on very few units is divided by a denominator much larger than its own exposure; the per-rule candidate counts above are the figures that describe each rule's own reliability.

Third, the savings figures are workload properties, not policy properties. Hash-identity savings range from 49.2\% (aiohttp) to 0.0\% (cobra, requests) and track the incidence of $A\rightarrow B\rightarrow A$ reversions rather than anything about the rule. A deployment should expect savings characteristic of its own revision behavior, and the aiohttp figure should not be read as a typical outcome.
\section{Discussion, Conclusion, and Future Work}\label{sec:discussion}

\subsection{Discussion}

\subsubsection{Informative Signals Are Not Necessarily Useful Ones}

The result we did not expect is that signals can be informative about retrieval impact,
be largely unassociated with one another, and still combine into a policy that changes no
decision. This is worth dwelling on, because the reasoning that led us to
attempt the combination is standard and, on its own terms, sound. We measured
per-signal discrimination, found that three of the four separated impacted from
unimpacted units while neighbourhood similarity did not, checked that the discriminating
signals were not redundant with one another, and concluded that combining them should
help. Every step of that argument is common practice in
software-analytics work. The conclusion was still wrong.

The gap is between two properties that correlational validation cannot distinguish. A
signal is \emph{informative} if knowing it reduces uncertainty about the outcome. A
signal is \emph{decision-relevant} if knowing it changes what a system does. Our results
separate these in two ways, and the second is the more general.

The first is redundancy. A signal that is individually informative can carry no
information the system does not already have. The substantive flag discriminates
impacted units in seven of ten repositories (Finding~5), yet it changes no re-embedding
decision at any budget a deployment would impose, because the history term already
reproduces its ordering (Results~3 and~5). Correlational validation cannot detect this, since it evaluates each signal
against the outcome rather than against the other signals in the presence of which it
will be used.

The second is objective mismatch, and it survives even when a signal is not redundant.
Retrieval-impact prediction and re-embedding scheduling are different problems: ranking
by $P(I>0)$ maximizes the number of impacted units re-embedded, whereas
$\mathrm{Hurt@B}$ counts the queries left visibly degraded, and the updates most likely
to be impacted are not the units whose staleness degrades the most queries. A signal validated against
the first can therefore misallocate a budget spent on the second.

We think the second pattern generalizes past our benchmark. Wherever a maintenance
mechanism spends a budget, the quantity being predicted and the quantity being optimized
tend to differ: predictive test selection estimates which tests are likely to fail,
while the objective is fault detection per unit of execution time; crawl scheduling
estimates which pages have changed, while the objective is the freshness a user
observes. A signal validated against the predicted quantity can be entirely correct and
still misallocate the budget. The methodological implication is that a maintenance
signal should be evaluated by the decision it produces under the objective it will face,
not by its association with a proxy for that objective. Such an evaluation needs an
artifact like CIPBench, which fixes the information boundary, the budget, and the metric
so that two policies can be compared on decisions rather than on scores.

\subsubsection{Why Filtering Succeeds Where Ranking Saturates}

The two tracks fail and succeed for opposite reasons, and the contrast is instructive.
Ranking saturates because the discriminative work is already done: history reproduces
the substantive ordering exactly, and exposure adds nothing on top. Filtering succeeds
because it
exploits a property no ranking signal captures---textual identity between the new state
and the indexed one. That property is not a better predictor of impact; it is a
\emph{proof} of non-impact. An update whose text is unchanged cannot move its
embedding, so it cannot change any ranking, and the miss rate is zero by construction
rather than by measurement.

This suggests a design principle for maintenance policies more broadly. Where a
decision admits a structural certificate---an argument that an action is safe rather
than an estimate that it is probably safe---that certificate is worth more than any
amount of signal engineering, and it should be sought first. Our largest single saving,
49.2\% in aiohttp, comes from a hash comparison. The four-signal weighted score that
occupied most of our design effort produces none.

\subsection{Conclusion}

This paper asks whether a changed code unit necessarily needs to be re-embedded. It
does not, but the useful answer turned out to lie in a different place from where we
looked for it.

Across 10 repositories in Python, Go, and JavaScript we find a consistent progression
from syntactic change to embedding displacement to retrieval impact, with links that
are informative rather than deterministic. CIPBench turns that observation into an
evaluation problem by requiring a policy to decide before observing the updated
embedding or its retrieval effect. Under that constraint, combining the available
pre-embedding signals into a ranking policy is statistically equivalent to using the
single strongest signal, for a structural reason we identify; conservative filtering,
by contrast, avoids real computation at zero measured risk.

The broader implication is that repository freshness need not be reduced to a choice
between refreshing everything and refreshing whatever changed most recently. A code
change can be treated as evidence whose importance depends on its syntax, its history,
and its position in the retrieval workload---but only where the maintenance mechanism
is actually sensitive to that evidence. Establishing where it is sensitive is what an
index-maintenance benchmark is for.

\subsection{Future Work}

\paragraph{Speculative re-embedding with read-time validation.}
Our policies decide once, at write time, and a wrong decision is silent: a unit skipped
in error stays stale until something happens to surface it. An alternative architecture
would treat the skip as a speculation and detect the misprediction afterwards, in the
manner of branch prediction. The obstacle is that naive detection is circular---
establishing whether a skip was wrong requires $I(u)$, which requires the embedding the
skip avoided---so the check costs exactly what it saves.

Finding~2 of Section~\ref{sec:empirical} suggests an escape. A stale unit that is never
retrieved cannot manifest its staleness, and most skipped units are never retrieved.
Validation could therefore move from write time to read time: retrieve $K'>K$ candidates
from the stale index, re-embed only the deferred units that surface within $K'$, and
re-rank before returning $K$. The cost is then bounded by $K'$ embeddings per query
rather than one per update, and is paid only for staleness that reaches a user. What
this cannot catch is a unit whose fresh embedding would enter the top-$K$ while its stale
embedding sits below $K'$; that residual miss rate is measurable from CIPBench's existing
labels as a function of $K'$, and bounding it---perhaps by per-change-class empirical
bounds on $\delta$, generalizing the exact certificate that hash identity already
provides---is the substantive open problem.

We regard this as the most promising direction, and as a systems problem rather than a
prediction problem: it needs a query workload with timing, a cost model in wall-clock
rather than embedding counts, and end-to-end retrieval quality under deferral.

\paragraph{Generality across languages and encoders.}
Extending CIPBench to Java, Rust, C++, and TypeScript would show how type systems and
language-specific conventions affect the observed relationships, and evaluating
code-specialized retrieval encoders would show whether the comment and string-literal
sensitivities persist when the representation model is designed for source code. A
harmonized cross-model impact pipeline would also permit a cleaner measurement of how
impact labels transfer between encoders than Section~\ref{sec:threats} can report.

\paragraph{Broader evolution and query workloads.}
Issue-linked streams give CIPBench a natural connection between repository evolution
and retrieval intent, but they are not production commit histories. Replication on
bot-filtered natural commit streams would show whether the temporal patterns and
repository-level differences persist under a less curated process. The impact
definition is likewise tied to ground-truth repository queries; question answering,
semantic code search, and completion-oriented retrieval expose different parts of a
repository and may therefore warrant different maintenance decisions. Developer intent
is a further unexploited signal: issue text is available but the current per-update
representation carries no validated intent label, and we preferred to exclude intent
rather than introduce a heuristic we could not evaluate under the same protocol.

\paragraph{Metric sensitivity and cumulative drift.}
The default $K=10$ and the binary impact predicate warrant comparison against graded
relevance measures such as NDCG change and against other values of $K$, and the
leave-one-repository-out grid was refined only near $w_1=0$. Separately, the present policies evaluate a unit once per refresh and carry nothing
between refreshes. A unit passed over at several consecutive refreshes accumulates
staleness that $\mathrm{Hurt@B}$ charges but that no term in Eq.~\ref{eq:score}
anticipates. A drift-aware variant would escalate a unit's priority
with its deferral count, restricted to units with retrieval exposure to lose; the
Track~A probing guard of Section~\ref{sec:tracka} already provides the sampling
machinery such a policy would need to calibrate itself.

\bibliographystyle{ACM-Reference-Format}
\bibliography{references}

\appendix
\renewcommand{\thesection}{Appendix \Alph{section}}
\begin{appendices}

\section{Signal-Independence Evidence}\label{app:signal-independence}

\begin{figure}[htbp]
\centering
\includegraphics[width=0.6\columnwidth]{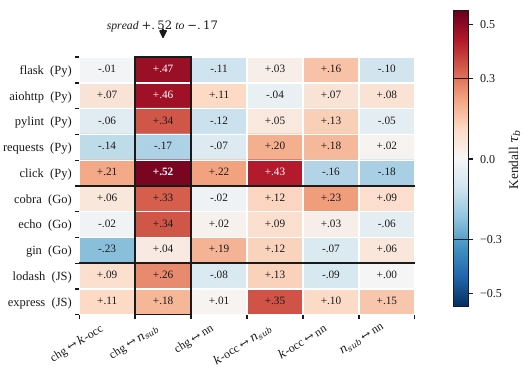}
\caption{Pairwise signal dependence across the 10 repositories. Fifty-two of the 60 associations fall below $|\tau|=0.3$. The boxed column is the exception discussed below: change type and $n_{sub}$ are moderately associated on average but range from $+0.52$ in click to $-0.17$ in requests, so the overlap between those two signals is itself repository-dependent.}
\label{fig:kendall-heatmap}
\end{figure}

Section~\ref{subsec:signals} reports how strongly the four inexpensive signals are
associated with one another. Table~\ref{tab:kendall-full} gives the complete 60-value
Kendall $\tau_b$ matrix and Figure~\ref{fig:kendall-heatmap} shows it. We treat
$|\tau|\ge0.5$ as a strong association, $0.3\le|\tau|<0.5$ as moderate, and
$|\tau|<0.3$ as weak.

\begin{table*}[htbp]
\caption{Pairwise Kendall $\tau_b$ among the four free signals across 10 repositories.}
\label{tab:kendall-full}
\centering
\scriptsize
\setlength{\tabcolsep}{3.7pt}
\begin{tabular}{llrrrrrr}
\toprule
\textbf{Repo} & \textbf{Lang.} & \textbf{change$\leftrightarrow k$-occ} & \textbf{change$\leftrightarrow n_{sub}$} & \textbf{change$\leftrightarrow$nn} & \textbf{$k$-occ$\leftrightarrow n_{sub}$} & \textbf{$k$-occ$\leftrightarrow$nn} & \textbf{$n_{sub}\leftrightarrow$nn}\\
\midrule
flask & Py & -.012 & +.475 & -.110 & +.028 & +.158 & -.101\\
aiohttp & Py & +.073 & +.460 & +.111 & -.036 & +.073 & +.076\\
pylint & Py & -.064 & +.340 & -.118 & +.052 & +.133 & -.051\\
requests & Py & -.138 & -.170 & -.073 & +.195 & +.179 & +.016\\
click & Py & +.210 & \textbf{+.516} & +.223 & +.433 & -.160 & -.177\\
cobra & Go & +.060 & +.327 & -.020 & +.119 & +.231 & +.092\\
echo & Go & -.016 & +.341 & +.021 & +.090 & +.029 & -.055\\
gin & Go & -.232 & +.045 & +.186 & +.121 & -.075 & +.056\\
lodash & JS & +.086 & +.262 & -.078 & +.127 & -.085 & +.001\\
express & JS & +.109 & +.183 & +.011 & +.348 & +.104 & +.147\\
\midrule
Mean $\tau$ & & +.008 & +.278 & +.015 & +.148 & +.059 & +.000\\
Mean $|\tau|$ & & .100 & .312 & .095 & .155 & .123 & .077\\
Repos with $|\tau|\ge.3$ & & 0 & 6 & 0 & 2 & 0 & 0\\
\bottomrule
\end{tabular}
\end{table*}

The matrix separates into one associated pair and five unassociated ones. Change type
versus $n_{sub}$ has mean $|\tau|=.312$ and reaches the moderate band or above in six of
the ten repositories; every other pair has mean $|\tau|\le.155$ and reaches it in at
most two. Across all 60 cells, 52 satisfy $|\tau|<0.3$.

The exception is not uniform, which is the more useful observation. Change type and
history are strongly associated in click ($+.516$) and flask ($+.475$), essentially
unassociated in gin ($+.045$), and \emph{negatively} associated in requests ($-.170$).
A definitional relationship between the two would not admit a negative value anywhere,
so the association is an empirical property of how these repositories evolve rather
than an artifact of how the signals are constructed. Where a repository accumulates long
runs of one change class, history and the current class agree; where evolution is more
heterogeneous, they diverge.

This is the empirical prior for Section~\ref{sec:trackb}. Retrieval exposure and
neighbourhood similarity carry information that neither change type nor history
contains, so combining them is motivated by measured complementarity rather than by
stacking redundant features. Change type, by contrast, duplicates history in most
repositories---which is consistent with the outcome reported in
Section~\ref{subsec:partition}, where cross-validation discards retrieval exposure and
the syntactic flag turns out to change no decision.

\section{Repository-Level Meta-Features}\label{app:meta-features}

Section~\ref{subsec:signals} reports that the four inexpensive signals differ in how
strongly they discriminate impacted units, and that the differences are
repository-dependent. This appendix records the associations we measured between
repository characteristics and per-signal strength. We report them for completeness
and as motivation for future work on applicability envelopes; no policy in this paper
consults them, and with only 10 repositories they do not support fine-grained routing
rules.

We ask whether repository characteristics are associated with the usefulness of individual signals. Table~\ref{tab:meta} reports Spearman correlations between five repository-level meta-features and per-repository signal strength.

\begin{table}[t]
\caption{Spearman rank correlation $\rho$ between repository meta-features and signal strength.}
\label{tab:meta}
\centering
\small
\begin{tabular}{lrrrr}
\toprule
\textbf{Meta-feature} &
\textbf{change\_type} &
\textbf{$k$-occ} &
\textbf{$n_{sub}$} &
\textbf{nn\_sim}\\
\midrule
size (number of units) & +0.406 & +0.248 & +0.564 & -0.382\\
temperature (substantive-update fraction) & +0.394 & -0.467 & +0.539 & -0.127\\
annotation density & -0.614 & -0.136 & -0.795$^*$ & +0.213\\
average history length & +0.503 & +0.091 & -0.006 & +0.285\\
$H$ fraction (impacted-unit fraction) & +0.018 & -0.164 & +0.358 & +0.418\\
\bottomrule
\end{tabular}
\vspace{2pt}
\begin{minipage}{0.96\columnwidth}\footnotesize
$^*p<0.05$. Larger $|\rho|$ indicates a stronger monotonic relationship.
\end{minipage}
\end{table}

The strongest observed association is between annotation density and the discrimination of $n_{sub}$ ($\rho=-0.795$): historical substantive-change count becomes less informative in heavily annotated repositories. This runs counter to the initial expectation that richer annotation histories would strengthen the usefulness of substantive-change history.

The aiohttp reversions from RQ1 provide one plausible explanation. In an annotation-heavy project, repeated type refinements can create a long history of syntactically substantive edits without necessarily producing persistent representation changes. Repository size is positively associated with $n_{sub}$ strength ($\rho=+0.564$), and grouping repositories by these meta-features does not reproduce clean programming-language boundaries. This again points toward evolution style rather than language alone.

We draw no finding from these correlations. Ten repositories cannot distinguish a
population-level relationship from a coincidence among the projects we happened to
select, and the one association that clears $p<0.05$ does so without a multiplicity
correction across the 20 tested pairs. The value of the table is that it identifies
annotation density and repository size as the two characteristics most worth
controlling for in a larger replication.

\section{Change-Type Persistence}\label{app:persistence}

Section~\ref{subsec:context} reports that substantive-change persistence varies more by
repository than by language. This appendix gives the underlying per-repository figures:
the rate at which consecutive updates to the same unit share a coarse change class, and
the dominant transition pattern in each repository.

\begin{table*}[t]
\caption{Change-type persistence and dominant transition patterns.}
\label{tab:persistence}
\centering
\small
\resizebox{\linewidth}{!}{
\begin{tabular}{llcl}
\toprule
\textbf{Repository} & \textbf{Lang.} &
\textbf{Persistence} &
\textbf{Most common transition pattern}\\
\midrule
aiohttp & Py & 95.9\% & signature $\rightarrow$ signature (69.8\%)\\
click & Py & 93.7\% & signature $\rightarrow$ signature (81.0\%)\\
pylint & Py & 94.6\% & implementation $\rightarrow$ implementation (43.5\%); comment $\rightarrow$ comment (33.2\%)\\
flask & Py & 65.0\% & mainly implementation $\rightarrow$ implementation; mixed evolution\\
requests & Py & 73.1\% & mainly implementation $\rightarrow$ implementation\\
cobra & Go & 69.8\% & implementation persistence mixed with implementation $\leftrightarrow$ signature\\
echo & Go & 66.4\% & implementation $\leftrightarrow$ signature (26.9\%)\\
gin & Go & 78.4\% & implementation $\leftrightarrow$ signature (15.0\%)\\
lodash & JS & 73.5\% & implementation persistence and implementation $\leftrightarrow$ signature\\
express & JS & 82.0\% & mainly implementation $\rightarrow$ implementation\\
\bottomrule
\end{tabular}
}
\end{table*}
The language-level averages are 84.5\% for Python, 71.5\% for Go, and 77.7\% for
JavaScript, but the within-language spread is larger than the between-language spread:
aiohttp, click, and pylint reach 93--96\%, whereas flask and requests fall to 65--73\%,
overlapping every Go and JavaScript repository. No policy in this paper consults these
values; we report them because they are the evidence that the persistence effect of
Finding~3 is not a language property.

\section{Robustness and Supporting Statistics}\label{app:robustness}

This appendix collects three sets of numbers that support claims in the main text
without being needed to follow them: the test statistics behind the RQ1 displacement
comparison, the language-level aggregation of the Track~B leaderboard, and the
substantive-weight sweep.

\subsection*{RQ1 displacement comparison}

Section~\ref{subsec:funnel} reports that the three change classes differ in embedding
displacement. A Kruskal--Wallis test over the 2{,}366 units gives $H=322.0$ with
$p=1.22\times10^{-70}$. All three pairwise Mann--Whitney comparisons have $p<10^{-9}$
and remain significant under the Bonferroni-adjusted threshold $\alpha'=0.017$, with
rank-biserial effect sizes between 0.39 and 0.68.

% CONFIRM AGAINST THE CODE: displacement is one value per unit, so this comparison is
% over units. If the released analysis instead pooled per-update step displacements,
% this paragraph and the matching one in T5 must be restated.
Because displacement is defined per unit (Section~\ref{subsec:empirical-setup}), each
unit enters this comparison exactly once and the groups are the three change classes.
A unit whose history spans several classes is assigned its most severe, so the class
label rather than the observation count is the source of dependence here: units with
heterogeneous histories are attributed to a class that describes only part of their
evolution. The non-independence caveat of Section~\ref{sec:threats} applies to the
per-update analyses, where a unit does contribute several observations.

\subsection*{Language-level Track~B results}

\begin{table}[htbp]
\caption{Mean $\Hurt@50$ by language, averaged over the repositories of each language.}
\label{tab:cipbench-lang}
\centering
\small
\begin{tabular}{lrrrrr}
\toprule
\textbf{Lang.} & Random & Recency & $k$-occ & YW & Oracle\\
\midrule
Python & .032 & .031 & .018 & .016 & .009\\
Go & .099 & .110 & .092 & \textbf{.051} & .025\\
JavaScript & .045 & .065 & .032 & .040 & .034\\
\bottomrule
\end{tabular}
\end{table}

Python is comparatively close to the oracle under these baselines, the Go repositories
leave the most room for improvement, and in the two JavaScript repositories $k$-occ has
lower mean harm than YW-hist. With three, three and two repositories per language
respectively, these means should be read as a compact summary of
Table~\ref{tab:cipbench-leaderboard} rather than as language-level estimates.

\subsection*{Substantive-weight sweep}

Section~\ref{subsec:w0} reports that the comparison between the weighted policy and
YW-hist is unchanged as the substantive coefficient $w_0$ varies. Table~\ref{tab:w0}
gives the sweep in full.

\begin{table}[htbp]
\caption{Mean $\Delta$ (Weighted $-$ YW-hist, pp) as a function of $w_0$. The per-repository weight selections are unchanged at every value of $w_0$, so the sweep isolates the substantive coefficient.}
\label{tab:w0}
\centering
\small
\begin{tabular}{crrr}
\toprule
$w_0$ & \textbf{Mean $\Delta$ (pp)} & \textbf{Median} & \textbf{Ties / 20}\\
\midrule
0 & +0.30 & 0 & 12\\
0.94 & +0.30 & 0 & 12\\
3.38 & +0.30 & 0 & 12\\
5.06 & +0.30 & 0 & 12\\
10 & +0.30 & 0 & 12\\
\bottomrule
\end{tabular}
\end{table}

Every summary statistic is identical at every value, and the per-repository weight
selections are unchanged throughout, so the sweep isolates the substantive coefficient
from the rest of the score. Section~\ref{subsec:partition} extends the check to the
full budget range, where the coefficient does alter the selection at six of 130 points.
\end{appendices}

\end{document}